\documentclass[11pt,a4paper]{article}
\pdfoutput=1
\usepackage{jheppub}
\usepackage{slashed}

\makeatletter
\def\@fpheader{\relax}
\makeatother

\usepackage[czech,english]{babel}
\usepackage{graphicx}
\usepackage{amsmath,amsfonts,amssymb,multirow,physics}
\usepackage{url}
\DeclareMathOperator{\MyProd}{\scalebox{1.4}{$\mathrm{I\kern-0.2ex I}$}}

\preprint{LCTP-24-08}

\title{Supersymmetric Charge Constraints on AdS Black Holes from Free Fields}

\author[a]{Finn Larsen}

\emailAdd{larsenf@umich.edu}

\author[a]{and Siyul Lee}

\emailAdd{siyullee@umich.edu}

\affiliation[a]{Leinweber Center for Theoretical Physics, University of Michigan, Ann Arbor, MI 48109, U.S.A.}

\abstract{
Supersymmetric AdS black hole solutions exist only when their angular momenta and charges satisfy a certain constraint that depends on the dimension. We show that these nonlinear relations on the conserved charges agree with a computation in the dual supersymmetric CFT in its free limit, with interactions entering only through a uniform rescaling of all charges. Our computations apply to the highly non-trivial charge constraints for AdS$_4$, AdS$_5$ and AdS$_7$ black holes, and generalize an earlier one for the analogous constraint in AdS$_3$. Our results suggest a microscopic understanding of AdS black holes beyond the scope of supersymmetric indices.}


\arxivnumber{}

\newcommand{\bea}{\begin{eqnarray}}
\newcommand{\eea}{\end{eqnarray}}
\newcommand{\la}{\label}
\newcommand{\be}{\begin{equation}}
\newcommand{\ee}{\end{equation}}

\newcommand{\nn}{\nonumber}

\begin{document}

\maketitle

\section{Introduction}\label{sec:introd}

Supersymmetric, asymptotically AdS black hole solutions to supergravity theories are known across various dimensions \cite{Cvetic:2005zi,Cassani:2019mms}, including
AdS$_3$ \cite{Banados:1992wn,Banados:1992gq}, AdS$_4$ \cite{Chong:2004na,Hristov:2019mqp},
AdS$_5$ \cite{Gutowski:2004ez,Gutowski:2004yv,Chong:2005da,Chong:2005hr, Kunduri:2006ek,Wu:2011gq}
and AdS$_7$ \cite{Chong:2004dy,Chow:2007ts,Wu:2011gp,Chow:2011fh,Bobev:2023bxl}.
They are parametrized by their conserved charges, i.e. the mass $E$ and a number of independent angular momenta $J_i$ and electric charges $Q_I$ that depend on the dimension. Because they are supersymmetric, they are all extremal. They have the lowest possible mass $E$ for the given charges, and therefore their temperature vanishes. However, in each dimension, regular supersymmetric black holes are known only when the remaining conserved charges $J_i$, $Q_I$ belong to a subset defined by a certain constraint. From the bulk perspective, these constraints on charges are sometimes linked to the absence of closed timelike curves. The goal of this article is to present a simple heuristic understanding of the constraints between the conserved charges in the dual SCFT. 

The supersymmetric, asymptotically AdS black holes are realized as ensembles of supersymmetric states/operators in the dual CFTs \cite{Maldacena:1997re,Gubser:1998bc,Witten:1998qj}. Over the last few years, there has been significant progress towards understanding the Bekenstein-Hawking entropy of the AdS black holes from the CFT side of this duality,
see \cite{Benini:2015eyy,Benini:2016rke,Hosseini:2017mds,Azzurli:2017kxo,
Cabo-Bizet:2018ehj,Choi:2018hmj,Benini:2018ywd,Choi:2018fdc,Choi:2019miv,
Choi:2019zpz,Nian:2019pxj,Hosseini:2019iad,Benini:2019dyp,Choi:2019dfu,
Cabo-Bizet:2020nkr,Larsen:2021wnu,Choi:2021rxi,Aharony:2024ntg}
among many others.
Nearly all of these advances rely on the supersymmetric index, a version of
the grand canonical partition function \cite{Kinney:2005ej,Romelsberger:2005eg,
Bhattacharya:2008zy}.\footnote{The study of black hole microstates
beyond the scope of the superconformal index has been very limited,
but recently there has been progress for small $N$ \cite{Chang:2022mjp,
Choi:2022caq,Choi:2023znd,Budzik:2023vtr,Choi:2023vdm}.}
In a grand canonical partition function,
the microstates are averaged with weights parametrized by chemical potentials.
Crucially, in the supersymmetric index the chemical potentials are restricted such that any microstate and its ${\cal Q}$-action, where ${\cal Q}$ is the supercharge preserved by the ensemble, contribute with \emph{equal magnitude} and \emph{opposite sign}. The resulting cancellations make the supersymmetric index independent of coupling, and so justify direct comparison of weakly-coupled CFT with the supergravity approximation to the bulk theory, which is dual to strongly-coupled CFT. The flip-side is that the supersymmetric index also becomes independent of a physical variable that is conjugate to the disallowed chemical potential.

The advance in recent years was to overcome the cancellations due to the \emph{opposite signs} of contributions from a microstate and its ${\cal Q}$-action. This was circumvented by introducing complexified chemical potentials that ensure a faithful count of the supersymmetric microstates, despite the cancellation between a supersymmetric microstate and its ${\cal Q}$-action. However, the \emph{equal magnitude} part has not been addressed. The supersymmetric index inevitably depends on one less chemical potentials than there are independent charges, so the index does not distinguish microstates along the direction in the space of conserved charges generated by the preserved supercharge. This fundamentally prevents the index from addressing the charge constraint, which contains information about the location in the space of charges.
Indeed, the charge constraint is surprising from the CFT side of the duality, because numerous local supersymmetric operators exist also for charges that violate the constraint.\footnote{Curiously, the charge constraint emerges in the microscopic accounting
of the supersymmetric black hole entropy from the condition that the extremum of the complex entropy function is real
\cite{Hosseini:2017mds,Choi:2018hmj,Choi:2018fdc}. This is suggestive, but given the intrinsic shortcoming of the index, it does not provide a satisfying microscopic explanation.}

In \cite{Larsen:2021wnu}, we made a proposal for the microscopic origin of the charge constraints for supersymmetric AdS$_3$ black holes. In any unitary supermultiplet of the small $\mathcal{N}=4$ super-Virasoro algebra, whether short or long, all weights appear in pairs. The two weights in each pair are separated in the charge configuration space along the direction of the preserved supercharge ${\cal Q}$, and the $R$-charges of the two weights average to $k$, the level of the $SU(2)_R$ algebra. This is precisely the condition that an extremal BTZ black hole is supersymmetric.
Therefore, in a supersymmetric ensemble of the dual CFT, defined as when the chemical potentials are such that all microstates along the direction of ${\cal Q}$ are weighted equally, the macroscopic charges of the ensemble as a whole satisfy the AdS$_3$ charge constraint. 

Generalization of this argument to higher dimensions is not straightforward. Superconformal algebras in higher dimensions are not as large and constraining as the super-Virasoro algebra in CFT$_2$, so they are consistent with more diverse multiplet structures. Moreover, in AdS$_{d+1}$ with $d>2$, the constraints on conserved charges that we want to illuminate are non-linear and highly non-trivial.

In this paper, we offer a heuristic derivation of the charge constraints.
In each dimension, we start from the free multiplet of the corresponding superconformal algebra. We then construct a grand canonical partition function that depends on as many chemical potentials as there are charges, thereby overcoming the fundamental limitation of the index. We define a supersymmetric ensemble that gives equal weight to all states along the direction generated by the supercharge, and compute the macroscopic charges of the ensemble. This procedure gives the correct functional form of the fully refined charge constraint in AdS$_5$, AdS$_4$, and AdS$_7$. The major heuristic element of our computation is the number of free multiplets in the theory, which we simply put in by hand. For example, for the $SU(N)$ SYM in $d=4$, we need $\frac{1}{2}N^2$ free multiplets, compared with $N^2$ in a genuinely free theory. This number sets the scale of all conserved charges. 

In the remainder of the introduction we discuss our prescription generically, without reference to any particular dimension. The body of the paper gives particulars for AdS$_3$, AdS$_5$, AdS$_4$ and AdS$_7$.

\subsection{Details of the Prescription}\label{sec:gendim}

AdS black holes are dual to ensembles of quantum states in a superconformal field theory in one fewer dimensions that all preserve the same amount of supersymmetry as the black holes. The local operators in the dual theory organize themselves into representations of the applicable superconformal algebra. Our starting point is the field content of the free representation, which provides the basic building blocks of the CFTs. It consists of free fields, both bosons and fermions, as well as derivatives that generate conformal descendants, and equations of motion that impose physical conditions. In a free CFT the particle number operator is well-defined, and so the free fields correspond to single particle states. There are infinitely many, because an arbitrary number of derivatives may act on the fields.

Every single particle state can be chosen as eigenstates of the Cartan generators of the bosonic subalgebra. The corresponding eigenvalues are the conformal dimension $E$, angular momenta $J_i$, and the R-symmetry charges $Q_I$, where the ranges of $i$ and $I$ depend on the dimension and on the amount of supersymmetry. The totals of the microscopic quantum numbers for the entire ensemble give the $E$, $J_i$, and $Q_I$ that we identify with the black hole charges. We only pick microscopic states that individually preserve the same supersymmetries as the black hole. These single particle BPS states are referred to as BPS letters. Since they are annihilated by the chosen supercharges, ${\cal Q} |\psi \rangle = 0$, the BPS states must saturate the unitarity bound $\{ {\cal Q}, {\cal Q}^\dag \} \geq 0$.
The superalgebra expresses the left hand side as a sum over the bosonic Cartan operators, so, schematically,
\bea\label{QQscheme}
\mathrm{BPS:} \quad \{ {\cal Q}, {\cal Q}^\dag \} = E - \sum_{i} J_i - \sum_{I} Q_I = 0~.
\eea
The quantum numbers of the BPS letters must satisfy the equality on the right, giving a linear BPS relation between the energy and the other conserved charges.\footnote{Depending on normalization of the charges, one or more terms in the sum \eqref{QQscheme} may contain numerical coefficients that differ from one.  An example is \eqref{7dunitbound} for the 6d $(2, 0)$ superconformal algebra. Throughout this paper, we use the notation of \cite{Cordova:2016emh}, to which we refer for details on the algebra and representations.}

The grand canonical partition function is the trace over all quantum states, with weights assigned to each state by chemical potentials that couple to the conserved charges. We define it with an explicit restriction to BPS states:
\bea\label{defZ}
Z &\equiv& \mathrm{Tr}_\mathrm{BPS} \left[ e^{-\beta \{ {\cal Q}, {\cal Q}^\dag \}}
e^{\sum_i \omega_i J_i + \sum_I \Delta_I Q_I} \right] \nn\\
&=& \mathrm{Tr}_\mathrm{BPS} \left[ e^{\sum_i \omega_i J_i + \sum_I \Delta_I Q_I} \right]~.
\eea
The second line is because the superalgebra \eqref{QQscheme} gives $\{ {\cal Q}, {\cal Q}^\dag \} = 0$. This removes the dependence on conformal dimension, but the partition function retains dependence on all chemical potentials $\omega_i$ and $\Delta_I$, there are as many of them as there are charges. Therefore, it is sensitive to the distribution of microstates along all directions in the charge space.

It is useful to define the grand canonical partition function over the BPS letters only. This gives the single particle BPS partition function
$Z_\mathrm{sp}$. However, in a quantum field theory, general states belong to a multiparticle Fock space that is generated by the single particle states in the usual way, with occupation numbers restricted by fermion or boson statistics. In the free theory any quantum number of a multiparticle state, including its energy, is the sum over the corresponding single particle quantum numbers. Therefore, the BPS partition function $Z$ over the entire BPS Hilbert space can be derived from the single particle BPS partition function $Z_\mathrm{sp}$, by taking combinatorics into account.

For example, for a single particle bosonic or fermionic BPS state that yields the single particle partition function $x_B$ or $x_F$, the partition function for the full Fock space is
\bea\label{countB}
1+x_B+x_B^2 + \cdots &=& \frac{1}{1-x_B}~, 
\eea
and $1 + x_F$, 
%
%
respectively. If there are $N_B$ bosonic and $N_F$ single particle BPS states,
each of which yields the single particle partition function $x_{B,i}$ and $x_{F,j}$,
the full partition function becomes
\begin{equation}
\label{countBF}
Z = \frac{\prod_{j=1}^{N_F} (1+x_{F,j})}{\prod_{i=1}^{N_B}(1-x_{B,i})}~. 
\end{equation}
%
%
These formulae are simply the standard Bose-Einstein and Fermi-Dirac distributions from elementary statistical physics, but expressed in a notation commonly used when discussing supersymmetric indices. In our prescription, we compute the multiparticle partition function as a simple exponential of the single particle partition function: 
\bea\label{assumeexp}
Z &=& e^{Z_\mathrm{sp}} = \exp\left(\sum_{i=1}^{N_B}x_{B,i} + \sum_{j=1}^{N_F}x_{F,j}\right)~. 
\eea
This is the limit of classical statistical physics. It is justified when the occupation number for any single particle state is so small that it is likely to be either $0$ or $1$. 
This assumption may be realized by the large number of gauge degrees of freedom for each single particle state. An improved treatment of such gauge degrees of freedom would project onto gauge singlets at the end, and that we do not do.

Given the grand canonical partition function $Z$ for the full Hilbert space, we can derive the macroscopic charges as ensemble averages in a standard manner. \eqref{defZ} gives
\bea\label{genericcharge}
Q_J &=& \frac{\mathrm{Tr}_\mathrm{BPS}
\left[ Q_J \cdot e^{\sum_i \omega_i J_i + \sum_I \Delta_I Q_I} \right]}
{\mathrm{Tr}_\mathrm{BPS}
\left[ e^{\sum_i \omega_i J_i + \sum_I \Delta_I Q_I} \right]}
= \pdv{\Delta_J} \log Z~, \nn\\
J_j &=& \frac{\mathrm{Tr}_\mathrm{BPS}
\left[ J_j \cdot e^{\sum_i \omega_i J_i + \sum_I \Delta_I Q_I} \right]}
{\mathrm{Tr}_\mathrm{BPS}
\left[ e^{\sum_i \omega_i J_i + \sum_I \Delta_I Q_I} \right]}
= \pdv{\omega_j} \log Z~.
\eea
These formulae express all the charges in terms of an equal number of chemical potentials.

\begin{table}
\begin{center}
\begin{tabular}{| c | c | c | c | c | c | c | c |}
\hline
$D$ & Charges & Constraint & SCA & Free multiplet & $N$ & $G_D$ 
\\
\hline
\hline
4 & $J,~Q_{1,2,3,4}$ & (\ref{4dcc}) & 3d $\mathcal{N}=8$ & $B_1 [0]^{[0,0,1,0]}_{1/2}$
& $\frac{\sqrt{2}}{3} N^{\frac32} = 1$ & $G_4 = \frac12$ 
\\
\hline
5 & $J_{1,2},~Q_{1,2,3}$ & (\ref{5dcc}) & 4d $\mathcal{N}=4$ & $B_1 \bar{B}_1 [0;0]^{[0,1,0]}_1$
& $\frac{1}{2}N^2=1$ & $G_5 = \frac{\pi}{4}$ 
\\
\hline
7 & $J_{1,2,3},~Q_{1,2}$ & (\ref{7dcc}) & 6d $(2, 0)$ & $D_1 [0,0,0]^{[1,0]}_2$
& $\frac{2}{3}N^3 =1$ & $G_7 = \frac{\pi^2}{8}$  
\\
\hline
\end{tabular}
\caption{\label{summary}
For each dimension of AdS$_D$, supersymmetric black hole solutions exist only when the indicated charges obey a nonlinear constraint. 
We derive a charge constraint with the identical form microscopically, from the free multiplet of the applicable superconformal algebra (SCA). The relative normalization of charges in the two computations corresponds to a numerical value for the number of free fields or, equivalently, for Newton's constant $G_D$. Further details are given in sections \ref{sec:5d}--\ref{sec:7d}.}
\end{center}
\end{table}

Denoting by ${\cal Q}$ the supercharge that is preserved by the black hole and by the dual BPS states, we now impose a linear relation between the chemical potentials such that quantum states that differ by the charges of ${\cal Q}$ are given the same weight. The statistical computations of macroscopic charges \eqref{genericcharge} were done prior to this stage, and included the gradient of the partition function in the direction along the constraint between the potentials. Therefore, the computation reflects the dependence of the partition function on all chemical potentials.

After the constraint on the chemical potentials is imposed, the statistical formulae \eqref{genericcharge} express all macroscopic charges in terms of one variable less than there are charges. Equivalently, the charges that are realized form a co-dimension one surface in the space of all charges, i.e. they satisfy a constraint. We find that the constraint on charges arrived at this way, from the field content of the microscopic theory, has the same highly non-trivial form as the non-linear charge constraint of the supersymmetric black holes. 

The black hole charges that satisfy the non-linear constraint in the gravitational theory, are in units of Newton's gravitational coupling constant. 
In contrast, in its simplest form, the microscopic computation considers a single free field. Our results are incomplete, because we do not determine the relative scale of the charges in the two computations. Comparison between the computations gives a value for Newton's constant or, equivalently, for the effective number of free fields. The summary in Table \ref{summary} records these values.

\subsection{Outline of the Article}\label{sec:outline}
In section \ref{sec:3d}, we review the proposal of \cite{Larsen:2021wnu} for the AdS$_3$ charge constraint. In CFT$_2$ there are no heuristic aspects, so we hope this makes the generic prescription more convincing. In sections \ref{sec:5d} through \ref{sec:7d}, we detail our heuristic derivation of the charge constraints in AdS$_5$, AdS$_4$ and AdS$_7$, respectively, following the prescription explained in the preceding subsection step by step. We conclude in section \ref{sec:discussion}, by discussing various shortcomings and implications of our arguments.


\section{AdS$_3$}\label{sec:3d}

In this section we discuss BTZ black holes in AdS$_3 \times S^3$ that are dual to 2d CFT with $(4, 4)$ superconformal symmetry.
The arguments of this section mostly appeared in section 5 of \cite{Larsen:2021wnu} and are included here as a simple template
for the prescription we apply in higher dimensions.

\subsection{The BTZ Black Hole and its Charge Constraint}\label{sec:3dBH}

The BTZ black holes in AdS$_3 \times S^3$ carry an energy $E$ and an angular momentum $J$ that both 
arise from the isometry of AdS$_3$, as well as two charges $Q_L$ and $Q_R$ associated with the isometry $SU(2)_L \times SU(2)_R$ of $S^3$.

One choice of $\frac14$-BPS sector in this theory corresponds to energy that saturates the unitarity bound:
\bea\label{3dunitbound}
E &\geq& J + Q_L - \frac{k_L}{2}~.
\eea
In this formula $k_L$ is the level of the $SU(2)_L$ current which, because of ${\cal N}=4$ supersymmetry, is related to the central charge as $c_L = 6k_L$. On the other hand, all black hole solutions in AdS$_3 \times S^3$ satisfy the extremality bound,
\bea\label{3dextbound}
E &\geq& J - \frac{Q_L^2}{2k_L}~,
\eea
which is saturated at vanishing temperature.
This formula is entirely gravitational, but we have simply expressed Newton's constant $G_3$
in terms of the level $k_L$ using the Brown-Henneaux formula for  
the central charge \cite{Brown:1986nw}.

A BTZ black hole can only be $\frac14$-BPS if it saturates both of (\ref{3dunitbound}) and (\ref{3dextbound}). That is only possible if
\bea\label{3dcc}
Q_L &=& k_L~.
\eea
This is the charge constraint on supersymmetric AdS$_3$ black holes. In a charge sector that violates \eqref{3dcc} there are no supersymmetric black holes. 

\subsection{Multiplets of CFT$_2$ with $(4, 4)$ Supersymmetry}

The dual 2d CFT has $(4, 4)$ supersymmetry. Its superconformal algebra factorizes into two independent copies of super-Virasoro algebra, and includes a bosonic subgroup $SO(2,2) \times SU(2)_L \times SU(2)_R$ that matches the isometry of AdS$_3 \times S^3$. To make progress, we first review representations of one chiral copy of the small $\mathcal{N}=4$ super-Virasoro algebra with $SU(2)$ R-symmetry \cite{Eguchi:1987sm,Eguchi:1987wf,Sevrin:1988ew,Eguchi:1988af}.

Each unitary representation of the super-Virasoro algebra is labeled by the $L_0$ and $Q_L$ eigenvalues $(h,q_L)$ of its superconformal primary. 
The entire multiplet consists of the primary and its descendants, each with its own weights $(h,q_L)$. We can focus on states that satisfy NS boundary conditions, because representations in the Ramond sector are isomorphic through spectral flow by a half-integral unit.

There are two qualitatively different types of representations: long multiplets whose primary has $h > q_L$ and
short multiplets whose primary has $h = q_L$.
The allowed values for $q_L$ of the primary are $0,1, \cdots, k_L-1$ for long multiplets,
and $0,1, \cdots, k_L$ for short multiplets.
The content of either type of representation can be summarized by its character defined by ${\rm Tr} ~q^{L_0} y^{Q_L}$, a function of two fugacities $q$ and $y$. The characters of the two types of representations are \cite{Eguchi:1987wf}:
\bea\label{3dN4char}
\text{Long : } ~{\rm ch}_{h,q_L}(q,y) &=& q^h F^{NS} 
\sum_{m=-\infty}^{\infty}  \left( y^{2(k_L+1)m +q_L+1}-y^{-2(k_L+1)m -q_L-1} \right) 
\frac{q^{(k_L+1)m^2+(q_L+1)m}}{y-y^{-1}} ~, \nn\\
\text{Short : } ~~~\chi_{q_L}(q,y) &=& q^{\frac{q_L}{2}} F^{NS}
\sum_{m=-\infty}^{\infty}  \left( \frac{y^{2(k_L+1)m +q_L+1}}
{(1+yq^{m+\frac12})^2}- \frac{y^{-2(k_L+1)m -q_L-1}}{(1+y^{-1}q^{m+\frac12})^2} \right)
\frac{q^{(k_L+1)m^2+(q_L+1)m}}{y-y^{-1}} ~, \nn\\
\eea
where
\bea\label{3dFNS}
F^{NS} = \prod_{n\geq1} \dfrac{\left( 1+yq^{n-\frac12} \right)^2 \left( 1+y^{-1}q^{n-\frac12} \right)^2 }{(1-y^2 q^n)(1-q^n)^2(1-y^{-2} q^n)} ~,
\eea
accounts for the action of creation operators, i.e. the negative frequency modes $\{ G_{r<0}\}$ and $\{ L_{n<0}, J^i_{n<0}\}$ of the four fermionic and four bosonic generators. 

\subsection{The Supersymmetric Ensemble}

The long and short multiplets discussed in the previous subsection are the only unitary representations of the super-Virasoro algebra.
Therefore, the microscopic duals of AdS$_3$ black holes introduced in section \ref{sec:3dBH} must organize themselves into these multiplets,
as well as the additional structure that would arise from the other copy of super-Virasoro algebra.

The long and short multiplet characters both exhibit the following property:
\bea\la{3dZ2even}
\chi_{q_L}(q,y) &=& \chi_{q_L}(q,q^{-1}y^{-1}) \cdot  q^{k_L} y^{2k_L} ~, \nn \\
{\rm ch}_{h,q_L}(q,y) &=& {\rm ch}_{h,q_L}(q,q^{-1}y^{-1}) \cdot  q^{k_L} y^{2k_L} ~.
\eea
This shows that, within any representation, a weight with $(h,q_L)$ is always paired with another one with $(h+k_L-q_L, 2k_L-q_L)$. The $Q_L$ eigenvalues of the two members of the pair are mirrored around $k_L$. Therefore, if both weights contribute equally to the grand canonical partition function, then the macroscopic charge $Q_L$ obtained as a statistical average over the ensemble will necessarily be $k_L$.

The condition that the weights $(h,q_L)$ and $(h+k_L-q_L, 2k_L-q_L)$ in the pair contribute equally to the grand canonical partition function is $yq^{1/2} = 1$. Therefore, when this condition is satisfied, the average $\langle Q_L \rangle =k_L$. Indeed, explicit computation shows that
\begin{equation}
\langle Q_L \rangle = y\frac{\partial \log Z}{\partial y}\Big|_{y=q^{-\frac{1}{2}}} = k_L~, 
\end{equation}
for any partition function that is a product of characters satisfying \eqref{3dZ2even}. We refer to grand canonical partition function with $yq^{1/2} = 1$ as the supersymmetric ensemble.  

Geometrically, the condition $yq^{1/2} = 1$ defining the supersymmetric ensemble means all quantum states along a straight line in the $(h,q_L)$ plane are counted equally. This is precisely the direction generated by the preserved supercharge, corresponding to one of the factors in the numerator of (\ref{3dFNS}). In other sections of this article, we generalize this definition to other dimensions. 

The definition of the supersymmetric ensemble is reminiscent of imposing $yq^{1/2} = -1$, the substitution that turns the grand canonical partition function into the index, or the elliptic genus. With the condition $yq^{1/2} = -1$, two microstates related by the supercharge ${\cal Q}$ contribute equal magnitude, but with opposite signs. Therefore, the only non-vanishing contributions are from short multiplets where the primary is annihilated by the supercharge. Moreover, combinations of short multiplets along the direction of the supercharge combine to a long multiplet
\begin{equation}
\chi_{h,q_L-1}(q,y) + 2\chi_{h,q_L}(q,y) + \chi_{h,q_L+1}(q,y)  = {\rm ch}_{h,q_L}(q,y)\Big|_{h=\frac{1}{2}q_L}~,
\end{equation}
and also cancel automatically in the index. With these cancellations, the index is unable to assign relative probabilities to the charges in this direction, and so it cannot account for the constraint. In contrast, the supersymmetric ensemble avoids massive cancellations and controls the direction generated by the supercharge by taking the average over all configurations. This prescription reproduces the charge constraint $\langle Q_L\rangle = k_L$ (\ref{3dcc}) that is satisfied for all supersymmetric black holes in AdS$_3 \times S^3$.

\section{AdS$_5$}\label{sec:5d}

In this section we discuss how the charge constraint of supersymmetric,
rotating and charged black holes in AdS$_5$,
emerges from its dual $\mathcal{N}=4$ Super-Yang-Mills theory in 4d.
We follow the prescription outlined in section \ref{sec:gendim}.

\subsection{The Black Hole and its Charge Constraint}\label{sec:5dBH}

Asymptotically AdS$_5$ black holes arise as solutions to type-IIB supergravity
in AdS$_5 \times S^5$ \cite{Gutowski:2004ez,Gutowski:2004yv,Chong:2005da,
Chong:2005hr,Kunduri:2006ek,Wu:2011gq}.
They carry the mass $E$ and two angular momenta $J_{1,2}$
for the isometry $SO(2,4)$ of AdS$_5$,
and three charges $Q_{1,2,3}$ for the isometry $SO(6)$ of $S^5$.
The black hole solution with all $6$ conserved quantities independent is known.

The black hole is supersymmetric when the unitarity bound between the mass and the charges
\bea\label{5dunitbound}
E \geq J_1 + J_2 + Q_1 + Q_2 + Q_3~,
\eea
is saturated. We have set the AdS$_5$ radius $\ell_5 = 1$.
Importantly, saturation is possible only when the charges obey an additional relation
\cite{Kim:2006he,Choi:2018hmj,Larsen:2019oll}
\bea\label{5dcc}
\left( Q_1Q_2Q_3 + \frac{N^2}{2}J_1J_2 \right)
- \left( Q_1+Q_2+Q_3 + \frac{N^2}{2}\right)
\left( Q_1Q_2 + Q_2Q_3 + Q_3Q_1 - \frac{N^2}{2}(J_1+J_2) \right) &=& 0~. \nn\\
\eea
We have traded the 5d Newton's constant $G_5$
into the field theory variable $N$ via 
$$
\frac{1}{2}N^2 = \frac{\pi\ell^3_5}{4G_5}~,
$$
for future convenience, but we stress that the origin of the charge constraint (\ref{5dcc})
is purely gravitational. (\ref{5dcc}) is the charge constraint for supersymmetric AdS$_5$ black holes.

We also present an unrefined ($J_1=J_2=J$ and $Q_1=Q_2=Q_3=Q$) version
of (\ref{5dcc}) that is more approachable, but still quite non-trivial:
\bea\label{5dccur}
\left( Q^3 + \frac{N^2}{2}J^2 \right)
- \left( 3Q + \frac{N^2}{2}\right)
\left( 3Q^2 - N^2J \right) &=& 0~. \nn\\
\eea

\subsection{The 4d $\mathcal{N}=4$ Free Vector Multiplet}\label{sec:5dmul}

The charged, rotating AdS$_5$ black holes introduced in the previous subsection
are dual to quantum states in the $\mathcal{N}=4$ Super-Yang-Mills theory in 4d.
In this subsection we introduce the free vector multiplet of the 4d $\mathcal{N}=4$ superconformal algebra 
$\mathfrak{psu}(2,2|4)$ that generates the single particle states.

Local operators can be organized into super-representations of the 4d $\mathcal{N}=4$ superconformal algebra. 
A super-representation consists of a superconformal primary and its descendants.
Following the notation of \cite{Cordova:2016emh}, we identify representations
by the Dynkin labels of the superconformal primary under the maximal bosonic subalgebra: 
$$
[j; \bar{j}]^{[R_1, R_2, R_3]}_E~.
$$
Here $E$ is the conformal weight, $j$, $\bar{j}$ are the integer-quantized Dynkin labels
for the $SU(2) \times SU(2)$ Lorentz group, and $R_{1,2,3}$ are the Dynkin labels
for the $SU(4)$ R-symmetry group.

The black hole charges used in section \ref{sec:5dBH} refer to the $SO(2,4) \times SO(6)$
isometry group of the AdS$_5 \times S^5$ geometry. They are charges of $SO(2)$ rotations in orthogonal $2$-planes.
The orthogonal basis are related to the Dynkin basis as: 
\bea
&& J_1 = \frac{j+\bar{j}}{2}~, \qquad \qquad ~~~~~~~
J_2 = \frac{j-\bar{j}}{2}~, \nn\\
&& Q_1 = R_2 + \frac{R_1+R_3}{2}~, \qquad
Q_2 = \frac{R_1+R_3}{2}~, \qquad
Q_3 = \frac{R_1-R_3}{2}~.
\eea
The energy $E$ is common to the two bases. We further note that $[R_1, R_2, R_3]$ are $SU(4)$ Dynkin labels, not to be confused with $SO(6)$ Dynkin labels that are related via $R_1 \leftrightarrow R_2$. In our conventions $[1,0,0]$ is {\bf 4} (fundamental of $SU(4)$ but spinor of $SO(6)$) and $[0,1,0]$ is {\bf 6} (fundamental of $SO(6)$ but antisymmetric tensor of $SU(4)$).

The supersymmetric black holes discussed in section \ref{sec:5dBH} preserve $\frac{1}{16}$ of the supersymmetry, so
they correspond to BPS states that are annihilated by $2$ out of $32$ Hermitian supercharges. We choose ${\cal Q}$ and ${\cal Q}^\dag$ that obey the algebra
\bea\label{5QSalgebra}
2\{{\cal Q},{\cal Q}^\dag\} &=&
E - \left( j + \frac32 R_1 + R_2 + \frac12 R_3 \right) \nn\\
&=& E-(Q_1+Q_2+Q_3+J_1+J_2) \geq 0~,
\eea
which plays the role of \eqref{QQscheme} in the generic prescription. As explained in section \ref{sec:gendim},
any field component can be identified with a weight in a representation, and so it is an eigenstate with 
respect to the operators $E$, $Q_I$ and $J_i$. It is BPS if the corresponding eigenvalues saturate \eqref{5QSalgebra}.

\begin{table}\catcode`\-=12
\begin{center}
\begin{tabular}{| c | c | c | c c c c c | c c c c c |}
\hline
& Bosonic Rep. & $E$ & $j$ & $\bar{j}$ & $R_1$ & $R_2$ & $R_3$ &
$J_1$ & $J_2$ & $Q_1$ & $Q_2$ & $Q_3$ \\
\hline
\hline
\multirow{9}{*}{Free fields} & \multirow{3}{*}{$[0;0]^{[0,1,0]}_1$} & 1 & 0 & 0 & 0 & 1 & 0 & 0 & 0 & 1 & 0 & 0 \\
& & 1 & 0 & 0 & 1 & $-1$ & 1 & 0 & 0 & 0 & 1 & 0 \\
& & 1 & 0 & 0 & 1 & 0 & $-1$ & 0 & 0 & 0 & 0 & 1 \\
\cline{2-13}
& \multirow{3}{*}{$[1;0]^{[0,0,1]}_\frac32$} & $\frac32$ & 1 & 0 & 0 & 0 & 1 &
$\frac12$ & $\frac12$ & $\frac12$ & $\frac12$ & $-\frac12$ \\
& & $\frac32$ & 1 & 0 & 0 & 1 & $-1$ &
$\frac12$ & $\frac12$ & $\frac12$ & $-\frac12$ & $\frac12$ \\
& & $\frac32$ & 1 & 0 & 1 & $-1$ & 0 &
$\frac12$ & $\frac12$ & $-\frac12$ & $\frac12$ & $\frac12$ \\
\cline{2-13}
& \multirow{2}{*}{$[0;1]^{[1,0,0]}_\frac32$} & $\frac32$ & 0 & 1 & 1 & 0 & 0 &
$\frac12$ & $-\frac12$ & $\frac12$ & $\frac12$ & $\frac12$ \\
& & $\frac32$ & 0 & $-1$ & 1 & 0 & 0 &
$-\frac12$ & $\frac12$ & $\frac12$ & $\frac12$ & $\frac12$ \\
\cline{2-13}
& $[2;0]^{[0,0,0]}_2$ & 2 & 2 & 0 & 0 & 0 & 0 &
1 & 1 &0 & 0 & 0 \\
\hline
\hline
Eq. of motion & $[1;0]^{[1,0,0]}_\frac52$ & $\frac52$ & 1 & 0 & 1 & 0 & 0 &
$\frac12$ & $\frac12$ & $\frac12$ & $\frac12$ & $\frac12$ \\
\hline
\hline
\multirow{2}{*}{Derivatives} & \multirow{2}{*}{$[1;1]^{[0,0,0]}_1$} & 1 & 1 & 1 & 0 & 0 & 0 & 1 & 0 & 0 & 0 & 0 \\
& & 1 & 1 & $-1$ & 0 & 0 & 0 & 0 & 1 & 0 & 0 & 0 \\
\hline
\end{tabular}
\caption{\label{5dfmtable} Generators of the BPS operators in the free vector multiplet
$B_1 \bar{B}_1 [0;0]^{[0,1,0]}_1$. The first $9$ rows are the components of free fields that satisfy the BPS condition. 
Below the first double horizontal lines is an equation of motion that effectively removes one of the $9$. The last two rows are derivatives that act on the $9-1=8$ operators, and so create $8$ towers of single particle operators.}
\end{center}
\end{table}

In 4d superconformal theories, a field $[j; \bar{j}]^{[R_1, R_2, R_3]}_E$ is a free field if at least one of $j$ and $\bar{j}$ is zero and, in addition, $E = 1+\frac{j+\bar{j}}{2}$.
There is one multiplet of the 4d $\mathcal{N}=4$ superconformal algebra that contains a free field: 
the free vector multiplet, $B_1 \bar{B}_1 [0;0]^{[0,1,0]}_1$. All that we need is Table \ref{5dfmtable}, where we summarize the BPS content of the free vector multiplet, i.e. all weights in the multiplet that saturate the unitarity bound (\ref{5QSalgebra}). 

There are $9$ field components that satisfy the BPS condition. The BPS bosons are $3$ of the $6$ scalars in the theory, 
and $1$ of the $2$ gauge field components. The fermions are, in the language of ${\cal N}=1$ supersymmetry, $3$ chiralini and $2$ gaugini. The entry below the first double line is an equation of motion that relates the two gaugini. It should be counted as a ``negative" field that serves to cancel some gaugini operators with derivatives acting on them. There are equations of motion for other free fields as well, but this component of the gaugino equation of motion is the only one that is consistent with the BPS condition.
The last two entries in Table \ref{5dfmtable} are derivatives that may act on any of the fields, and on the equation of motion, to produce BPS descendants. The gradient operator has $4$ components in $4$ dimensions, but only $2$ preserve the BPS-ness of the field. The $9-1=8$ free fields and their derivatives generate the entire list of supersymmetric operators in the free vector multiplet. From a bulk point of view, these are the single particle BPS states.

\subsection{The Supersymmetric Ensemble}

Given the exhaustive list of single particle BPS states generated by the supersymmetric operators in Table \ref{5dfmtable},
we can now define a grand canonical partition function $Z_\mathrm{sp}$ over the single particle states. Rather than the chemical potentials as in \eqref{defZ}, we use fugacities $(p,q,x,y,z)$ that are related by
\bea
e^{\omega_1} = p^2~, \qquad
e^{\omega_2} = q^2~, \qquad
e^{\Delta_1} = x^2~, \qquad
e^{\Delta_2} = y^2~, \qquad
e^{\Delta_3} = z^2~, 
\eea
and so define the single particle BPS partition function by
\bea\label{5ddeff}
Z_\mathrm{sp} &\equiv&
\mathrm{Tr}_\mathrm{BPS} \left[ p^{2J_1} q^{2J_2} x^{2Q_1} y^{2Q_2} z^{2Q_3} \right]~.
\eea
The maneuver doubling the exponents avoids fractional powers, although the subtle feature of non-analyticity 
and ``second sheet'' \cite{Cassani:2021fyv} is not relevant to our purpose.

We read off the single particle partition function $Z_\mathrm{sp}$ from Table \ref{5dfmtable}.
The sum over the weights of the $8=9-1$ free fields gives
\bea
x^2+y^2+z^2+xyzpq
\left(\frac{1}{x^2} + \frac{1}{y^2} + \frac{1}{z^2} + \frac{1}{p^2} + \frac{1}{q^2} -1 \right)
+ p^2q^2~.
\eea
Any number of the two derivatives that preserve the BPS condition can act on each of the free fields, and on the equation of motion. Each derivative contributes a factor of $p^2$ or $q^2$, so we need a geometric sum over these. We then find the single particle BPS partition function
\bea\label{5dsppf}
Z_\mathrm{sp} &=& \frac{x^2+y^2+z^2+xyzpq
\left(\frac{1}{x^2} + \frac{1}{y^2} + \frac{1}{z^2} + \frac{1}{p^2} + \frac{1}{q^2} - 1\right)
+ p^2q^2}{(1-p^2)(1-q^2)}~.
\eea
According to our prescription discussed in section \ref{sec:gendim}, the full partition function is equal to the exponential of the single particle partition function \eqref{5dsppf}:
\bea\label{5dlogZ}
Z &\equiv& \exp[ Z_\mathrm{sp}]~.
\eea

From this grand canonical partition function, we obtain the macroscopic charges as ensemble averages in the standard manner. Changing variables $(\Delta_I, \omega_i) \, \to \, (p,q,x,y,z)$, \eqref{genericcharge} becomes
\bea
2Q_1 = x \pdv{x} \log Z~, \qquad 2J_1 = p \pdv{p} \log Z~,
\eea
and analogously for the charges with different indices.
The charges obtained from (\ref{5dlogZ}) in this way  are
\bea\label{5dchgpotrel}
Q_1 &=& \frac{2x^2+xyzpq
\left(-\frac{1}{x^2} + \frac{1}{y^2} + \frac{1}{z^2} + \frac{1}{p^2} + \frac{1}{q^2} - 1\right)
}{2(1-p^2)(1-q^2)}~, \\
J_1 &=& \frac{2p^2(q^2+x^2+y^2+z^2) + xyzpq(1+p^2)\left(
\frac{1}{x^2} + \frac{1}{y^2} + \frac{1}{z^2} - \frac{1}{p^2} + \frac{1}{q^2} - 1\right)
+4xyzpq }{2(1-p^2)^2(1-q^2)}~, \nn
\eea
and similarly for the permutations. (\ref{5dchgpotrel}) express the $5$ average charges of the ensemble in terms of $5$ potentials. 
BPS states are populated throughout the five-dimensional charge space, not just on some specific hypersurface thereof.
Thus, the $5$ average charges may take generic values without any particular constraint as well,
as the $5$ potentials are varied.

We now define the supersymmetric ensemble as a grand canonical ensemble where the operators that are separated in the charge space along the direction of the preserved supercharge are weighed equally. The preserved supercharge ${\cal Q}$ carries quantum numbers
$(E, J_1, J_2, Q_1, Q_2, Q_3) =
\left( \frac12, -\frac12, -\frac12, \frac12, \frac12, \frac12 \right)$, so the supersymmetric ensemble corresponds to the relation
\bea\label{5dsusyens}
\frac{xyz}{pq} &=& 1~,
\eea
between the fugacities.
For the supersymmetric ensemble satisfying (\ref{5dsusyens}), there is one relation between the five charges (\ref{5dchgpotrel}):
\bea\label{5dccmicro}
\left( Q_1Q_2Q_3 + J_1J_2 \right)
- \left( Q_1+Q_2+Q_3 + 1\right)
\left( Q_1Q_2 + Q_2Q_3 + Q_3Q_1 - J_1-J_2 \right) &=& 0~. ~~~~~~~
\eea
This is precisely the supersymmetric AdS$_5$ black hole charge constraint (\ref{5dcc}) with
\bea\label{5dNvalue}
\frac{1}{2}N^2 = 1 \quad \leftrightarrow \quad  \frac{\pi\ell^3_5}{4G_5} = 1~.
\eea
Equivalently, the statistical constraint \eqref{5dccmicro} agrees with the
macroscopic constraint (\ref{5dcc}) if macroscopic charges are in units of $\frac{1}{2}N^2$. 
A truly free $SU(N)$ theory would have $N^2$ identical copies of the free fields.
We interpret the remaining relative factor $\frac{1}{2}$ 
as a reduction that is due to interactions, but we claim no quantitative understanding of this factor. This feature non-withstanding, our computation establishes the functional dependence on charges of the constraint (\ref{5dcc}) from the combinatorics of free fields. 

The unrefined charges are defined by taking $x=y=z$ and $p=q$ in (\ref{5dchgpotrel}):
\bea\label{5dchgpotrelur}
Q &=& \frac{2x^2 + 2 x^3 + xp^2 - x^3p^2}{2(1-p^2)^2}~, \nn\\
J &=& \frac{3x(1+x)^2p^2 + (2+3x-x^3)p^4}{2(1-p^2)^3}~. 
\eea
It follows automatically that, upon picking the supersymmetric ensemble $x^3=p^2$, 
these charges satisfy the unrefined charge constraint (\ref{5dccur}) with $\frac{1}{2}N^2 = 1$.

\section{AdS$_4$}\label{sec:4d}

In this section we derive the charge constraint for the supersymmetric AdS$_4$ black holes.
The AdS$_4$ theory and its dual CFT$_3$ have features that are absent in AdS$_5$/CFT$_4$, such as magnetic charges and
the Chern-Simons term. Such complications are not directly relevant to our computation. We
find the charge constraint of the supersymmetric, rotating and electrically charged black holes in AdS$_4$
from the free hypermultiplet of the 3d $\mathcal{N}=8$ superconformal algebra.

\subsection{The Black Hole and its Charge Constraint}\label{sec:4dBH}

Asymptotically AdS$_4$ black holes arise as solutions to the
4d gauged supergravity theories \cite{Chong:2004na,Hristov:2019mqp}.
They carry the mass $E$ and an angular momentum $J$ for the
isometry $SO(2,3)$ of AdS$_4$,
and four electric charges $Q_{1,2,3,4}$ for the isometry $SO(8)$ of $S^7$.
The solution with the four electric charges pairwise equal ($Q_1=Q_3$ and $Q_2=Q_4$) was found in \cite{Chong:2004na}, 
and the most general solution with all four electric charges independent was found in \cite{Hristov:2019mqp}.

The black hole is supersymmetric when the unitarity bound between
the mass and the charges
\bea\label{4dunitbound}
E \geq J + \frac12 \left( Q_1 + Q_2 + Q_3 + Q_4 \right)~,
\eea
is saturated. However, the saturation is possible only when the charges obey the additional relation
\cite{Choi:2018fdc,Hristov:2019mqp}\footnote{Although the solution with all four electric charges
independent was found in \cite{Hristov:2019mqp}, its charge constraint had been correctly
conjectured earlier \cite{Choi:2018fdc}, based on the solution with pairwise equal charges \cite{Chong:2004na} and the structure of the entropy function.}
\bea\label{4dcc}
(\mathbb{Q}_3)^2 - (\mathbb{Q}_1) (\mathbb{Q}_2) (\mathbb{Q}_3)
+ (\mathbb{Q}_1)^2 (\mathbb{Q}_4) &=& 0~,
\eea
where we used the shorthand notation
\bea
(\mathbb{Q}_1) &\equiv& Q_1+Q_2+Q_3+Q_4~, \nn\\
(\mathbb{Q}_2) &\equiv& Q_1Q_2+Q_1Q_3+Q_1Q_4 + Q_2Q_3 + Q_2Q_4 + Q_3Q_4 + \frac{2N^3}{9}~, \nn\\
(\mathbb{Q}_3) &\equiv& Q_1Q_2Q_3 + Q_1Q_2Q_4 + Q_1Q_3Q_4 + Q_2Q_3Q_4 - \frac{4N^3}{9}J~, \nn\\
(\mathbb{Q}_4) &\equiv& Q_1Q_2Q_3Q_4 + \frac{2N^3}{9} J^2~.
\eea
In the formulae above, we set the AdS$_4$ radius $\ell_4 = 1$. We
traded the 4d Newton's constant for the field theory variable $N$
via $$N^{\frac32} = \frac{3}{2\sqrt{2}G_4}~,$$
for future convenience, but we stress that the origin of the charge constraint (\ref{4dcc})
is purely gravitational. (\ref{4dcc}) is the supersymmetric AdS$_4$ black hole charge constraint.

To make the formulae more approachable and to make the connection to the literature,
we also present the charge constraint (\ref{4dcc}) with pairwise equal electric charges
(see e.g. \cite{Choi:2018fdc})
\bea\label{4dccurp}
Q_1Q_2 (Q_1+Q_2)^2 - (Q_1+Q_2) \cdot \frac{2N^3}{9} J - \frac{2N^3}{9} J^2 &=& 0~.
\eea
as well as the version with all four electric charges equal (see e.g. \cite{Larsen:2020lhg}):
\bea\label{4dccur}
4 Q^4 - 2Q \cdot \frac{2N^3}{9} J - \frac{2N^3}{9} J^2 &=& 0~.
\eea
The formulae simplify greatly, but they remain quite nontrivial. 

\subsection{The 3d $\mathcal{N}=8$ Free Hypermultiplet}\label{sec:4dmul}

In this subsection we present the free hypermultiplet of the 3d $\mathcal{N}=8$
superconformal algebra, from which the AdS$_4$ charge constraint (\ref{4dcc}) will be derived in the next subsection.

The 3d $\mathcal{N}=8$ superconformal algebra has maximal bosonic subalgebra
$\mathfrak{so}(2,3) \oplus \mathfrak{so}(8)$, matching the isometry of AdS$_4 \times S^7$.
Local operators in the theory are organized into representations of this subalgebra.
A super-representation of the 3d $\mathcal{N}=8$ superconformal algebra is uniquely specified by 
the Dynkin labels of its superconformal primary.
Following the notation of \cite{Cordova:2016emh}, we write representations of the
bosonic subalgebra as
$$[j]^{[R_1, R_2, R_3, R_4]}_E~,$$
where $E$ is the conformal weight, $j$ is the integer-quantized $SO(3)$ Dynkin label,
and $[R_1,R_2, R_3, R_4]$ are the $SO(8)$ Dynkin labels so that $[1,0,0,0]$ is the vector {\bf 8}.

The black hole charges used in section \ref{sec:4dBH} refer to the
orthogonal basis that is related to the Dynkin basis as
\bea
&& J= \frac{j}{2} ~, \\
&& Q_1 = R_3 + R_2 + \frac{R_1+R_4}{2}~, \quad
Q_2 = R_2 + \frac{R_1+R_4}{2}~, \quad
Q_3 = \frac{R_1+R_4}{2}~, \quad
Q_4 = \frac{R_1-R_4}{2}~. \nn
\eea
This relation between the orthogonal and the Dynkin bases of $SO(8)$ differs from the more conventional one by $R_1 \leftrightarrow R_3$. We have exploited the $S_3$ outer automorphism of $SO(8)$ to match the convention (\ref{4dunitbound}) with that of \cite{Cordova:2016emh}.

A (not necessarily the highest) weight $[j]^{[R_1, R_2, R_3, R_4]}_E$
is annihilated by our choice of supercharge ${\cal Q}$ if it saturates the unitarity bound
\bea\label{4dunitboundmicro}
E &\geq& \frac12 j + R_1 + R_2 + \frac12 R_3 + \frac12 R_4 \nn\\
&=& J + \frac12 \left( Q_1 + Q_2 + Q_3 + Q_4 \right)~,
\eea
that every weight must satisfy. Such weights correspond to local BPS operators.

\begin{table}\catcode`\-=12
\begin{center}
\begin{tabular}{| c | c | c | c c c c c | c c c c c |}
\hline
& Bosonic Rep. & $E$ & $j$ & $R_1$ & $R_2$ & $R_3$ & $R_4$ &
$J$ & $Q_1$ & $Q_2$ & $Q_3$ & $Q_4$ \\
\hline
\hline
\multirow{8}{*}{Free fields} & \multirow{4}{*}{$[0]^{[0,0,1,0]}_{\frac12} $} & $\frac12$ & 0 & 0 & 0 & 1 & 0 & 0 & 1 & 0 & 0 & 0 \\
& & $\frac12$ & 0 & 0 & 1 & $-1$ & 0 & 0 & 0 & 1 & 0 & 0 \\
& & $\frac12$ & 0 & 1 & $-1$ & 0 & 1 & 0 & 0 & 0 & 1 & 0 \\
& & $\frac12$ & 0 & 1 & 0 & 0 & $-1$ & 0 & 0 & 0 & 0 & 1 \\
\cline{2-13}
& \multirow{4}{*}{$[1]^{[0,0,0,1]}_1$} & 1 & 1 & 0 & 0 & 0 & 1 &
$\frac12$ & $\frac12$ & $\frac12$ & $\frac12$ & $-\frac12$ \\
& & 1 & 1 & 0 & 1 & 0 & $-1$ &
$\frac12$ & $\frac12$ & $\frac12$ & $-\frac12$ & $\frac12$ \\
& & 1 & 1 & 1 & $-1$ & 1 & 0 &
$\frac12$ & $\frac12$ & $-\frac12$ & $\frac12$ & $\frac12$ \\
& & 1 & 1 & 1 & 0 & $-1$ & 0 &
$\frac12$ & $-\frac12$ & $\frac12$ & $\frac12$ & $\frac12$ \\
\hline
\hline
Derivative & $[2]^{[0,0,0,0]}_1$ & 1 & 2 & 0 & 0 & 0 & 0 & 1 & 0 & 0 & 0 & 0 \\
\hline
\end{tabular}
\caption{\label{4dfmtable}
Components of the BPS operators in the free hypermultiplet $B_1 [0]^{[0,0,1,0]}_{\frac12}$.
The first $8$ rows are free fields. Below the double horizontal lines is the one derivative that preserves BPS.}
\end{center}
\end{table}

In 3d superconformal theories, a field $[j]^{[R_1, R_2, R_3, R_4]}_E$
is free if $j \leq 1$ and $E = \frac{j+1}{2}$.
There are two multiplets of the 3d $\mathcal{N}=8$ superconformal algebra
that contain a free field \cite{Cordova:2016emh}.
The free hypermultiplets $B_1 [0]^{[0,0,1,0]}_{\frac12}$
and $B_1 [0]^{[0,0,0,1]}_{\frac12}$ are related by a $Z_2$ subgroup of the outer automorphism of $SO(8)$,
so we can choose $B_1 [0]^{[0,0,1,0]}_{\frac12}$ without loss of generality.
The rest of this section would be reproduced with minimal relabeling
had we chosen otherwise.

In Table \ref{4dfmtable} we summarize all weights in this free hypermultiplet that
saturate the unitarity bound (\ref{4dunitboundmicro}). There are $8$ free fields: $4$ scalars and $4$ spinors.
There is no equation of motion that is compatible with the BPS condition. The last entry is a derivative that can act on any of the fields and so produce its BPS descendants.
Note that out of 3 derivatives in 3 dimensions, only 1 preserves the BPS-ness of the field. 
The 8 free fields and their derivatives
are the exhaustive list of supersymmetric operators in the free hypermultiplet.

\subsection{The Supersymmetric Ensemble}

We now compute the single particle BPS partition function as a trace over the free hypermultiplet states given in Table \ref{4dfmtable}, with fugacities $(p,x,y,z,w)$ conjugate to each charge:
\bea
Z_{\rm sp} &\equiv& {\rm Tr}_{\rm BPS} \left[ p^{2J} x^{2Q_1} y^{2Q_2} z^{2Q_3} w^{2Q_4} \right]\nn\\
%
%
\label{4dsppf}
&=& \frac{x^2+y^2+z^2+w^2+ pxyzw \left( \frac{1}{x^2} + \frac{1}{y^2} + \frac{1}{z^2}
+ \frac{1}{w^2} \right) }{1-p^2}~.
\eea
It is the derivative that gives rise to the geometric series in $p^2$.
(\ref{4dsppf}) is the single particle partition function.

The grand canonical partition function over the full Hilbert space
is given by the ordinary exponential of the single particle partition function:
$Z \equiv \exp[Z_{\rm sp}]$. We then compute the macroscopic charges as statistical averages. They are
\bea\label{4dchgpotrel}
Q_1 &=& \frac{pxyzw \left( \frac{1}{x^2} + \frac{1}{y^2} + \frac{1}{z^2}
+ \frac{1}{w^2} \right) + 2x^2 - \frac{2 p xyzw}{x^2} }{2 (1-p^2)} ~, \\
J &=& \frac{pxyzw \left( \frac{1}{x^2} + \frac{1}{y^2} + \frac{1}{z^2}
+ \frac{1}{w^2} \right)}{2 \left(1-p^2\right)} + \frac{p^2 \left( x^2+y^2+z^2+w^2
+ p xyzw \left(\frac{1}{x^2}+\frac{1}{y^2}+\frac{1}{z^2}+\frac{1}{w^2}\right)\right)}{\left(1-p^2\right)^2}~.  \nn
\eea
Analogous expressions for $Q_2$, $Q_3$, and $Q_4$ follow by simple permutations of indices. 

Finally, we define the supersymmetric ensemble as a grand canonical ensemble
where the operators that are separated in the charge space along the direction
of the preserved supercharge are weighed equally.
The preserved supercharge ${\cal Q}$ carries quantum numbers
$(E, J, Q_1, Q_2, Q_3, Q_4) =
\left( \frac12, -\frac12, \frac12, \frac12, \frac12, \frac12 \right)$,
so the supersymmetric ensemble corresponds to imposing the relation
\bea\label{4dsusyens}
\frac{xyzw}{p} &=& 1~,
\eea
between the fugacities.

Because of the relation \eqref{4dsusyens} between the $5$ potentials, in the supersymmetric ensemble 
the $5$ charges $Q_{1, 2, 3, 4}$ and $J$ are not independent. The expressions (\ref{4dchgpotrel})
give the relation:
\bea\label{4dccmicro}
(\mathbb{Q}_3)_{\frac92}^2
- (\mathbb{Q}_1)_{\frac92} (\mathbb{Q}_2)_{\frac92} (\mathbb{Q}_3)_{\frac92}
+ (\mathbb{Q}_1)_{\frac92}^2 (\mathbb{Q}_4)_{\frac92} &=& 0~,
\eea
where
\bea
(\mathbb{Q}_1)_{\frac92}
&\equiv& Q_1+Q_2+Q_3+Q_4~, \nn\\
(\mathbb{Q}_2)_{\frac92} 
&\equiv& Q_1Q_2+Q_1Q_3+Q_1Q_4 + Q_2Q_3 + Q_2Q_4 + Q_3Q_4 + 1 ~, \nn\\
(\mathbb{Q}_3)_{\frac92}
&\equiv& Q_1Q_2Q_3 + Q_1Q_2Q_4 + Q_1Q_3Q_4 + Q_2Q_3Q_4 - 2J~, \nn\\
(\mathbb{Q}_4)_{\frac92}
&\equiv& Q_1Q_2Q_3Q_4 + J^2~.
\eea
It is precisely the supersymmetric AdS$_4$ black hole charge constraint (\ref{4dcc}) with the numerical values
\bea\label{4dNvalue}
 \frac{\sqrt{2}}{3} N^{\frac32} = 1 \quad \leftrightarrow \quad G_4 = \frac12~. 
\eea
We interpret this relative scale of all charges as the effective number of free multiplets needed to account for the constraint. 
%
%

The formulae simplify significantly when we do not distinguish between all $4$ electric charges. 
First, let $z=x$ and $w=y$ in (\ref{4dchgpotrel}):
\bea\label{4dchgpotrelurp}
Q_1 = Q_3 &=& \frac{x^2}{1-p} ~, \nn\\
Q_2 = Q_4 &=& \frac{y^2}{1-p} ~, \nn\\
J &=& \frac{p (x^2+y^2)}{(1-p)^2}~.
\eea
The definition of the supersymmetric ensemble \eqref{4dsusyens} simplifies to $p=x^2y^2$, 
and then the charges \eqref{4dchgpotrelurp} satisfy
\bea\label{4dccmicrourp}
Q_1Q_2 (Q_1+Q_2)^2 - (Q_1+Q_2) J - J^2 &=& 0~. 
\eea
This is the pairwise unrefined version of the charge constraint (\ref{4dccurp}) with
$\frac{\sqrt{2}}{3} N^{\frac32} = 1$.

To treat all $4$ electric charges as identical, we further let $x=y$ in (\ref{4dchgpotrelurp}):
\bea\label{4dchgpotrelur}
Q \equiv Q_{1,2,3,4} &=& \frac{x^2}{1-p} ~, \nn\\
J &=& \frac{2 p x^2}{(1-p)^2}~.
\eea
These charges, with the equation $p=x^4$ defining the supersymmetric ensemble, satisfy
\bea\label{4dccmicrour}
4 Q^4 - 2QJ - J^2 &=& 0~. 
\eea
This is the fully unrefined version of the charge constraint (\ref{4dccur}) with 
$\frac{\sqrt{2}}{3} N^{\frac32} = 1$.

\section{AdS$_7$}\label{sec:7d}

In this section we derive the charge constraint for the supersymmetric,
rotating and charged black holes in AdS$_7$.
from the dual $(2, 0)$ theory in 6d.

\subsection{The Black Hole and its Charge Constraint}\label{sec:7dBH}

Asymptotically AdS$_7$ black holes arise as solutions to a consistent truncation
of the 11d supergravity on $S^4$.
They carry the mass $E$ and
three angular momenta $J_{1,2,3}$ for the isometry $SO(2,6)$ of AdS$_7$,
and two charges $Q_{1,2}$ for the isometry $SO(5) \sim Sp(4)$ of $S^4$.
Particular solutions with equal angular momenta \cite{Chong:2004dy},
those with equal charges \cite{Chow:2007ts}
and those with two vanishing angular momenta and two independent charges
\cite{Wu:2011gp,Chow:2011fh} were constructed some time ago,
but the solution with all angular momenta and charges independent
was found only recently in \cite{Bobev:2023bxl}.

These black holes are supersymmetric when the unitarity bound between the
mass and the charges
\bea\label{7dunitbound}
E \geq J_1 + J_2 + J_3 + 2Q_1 + 2Q_2~,
\eea
is saturated.
However, the saturation is possible only when the charges obey the
additional relation \cite{Larsen:2020lhg,Bobev:2023bxl}\footnote{
The convention for charges differ from that of \cite{Bobev:2023bxl}
by $J_i^{\rm here} = J_i^{\rm there}$ and $Q_i^{\rm here} = \frac{Q_i^{\rm there}}{2}$.}
\bea\label{7dcc}
&& \frac12 (Q_1^2 + Q_2^2 + 4Q_1Q_2) + \frac{N^3}{3} (J_1+J_2+J_3)
- \frac{Q_1Q_2(Q_1+Q_2) - \frac{N^3}{3} (J_1J_2 + J_2J_3 + J_3J_1)}
{Q_1+Q_2- \frac{N^3}{3}} \nn\\
&=& \sqrt{
\left( \frac12 (Q_1^2 + Q_2^2 + 4Q_1Q_2) + \frac{N^3}{3} (J_1+J_2+J_3) \right)^2
- \left( Q_1^2Q_2^2 + \frac{2N^3}{3}J_1J_2J_3 \right)}~.
\eea
In the formulae above, we set the AdS$_7$ radius $\ell_7 = 1$.
We traded the 7d Newton's constant for the field theory variable $N$ via
$$N^3 = \frac{3\pi^2}{16G_7}~,$$
for future convenience,
but we stress that the origin of the charge constraint (\ref{7dcc}) is purely gravitational.
(\ref{7dcc}) is the supersymmetric AdS$_7$ black hole charge constraint.
We also present an unrefined ($J_1=J_2=J_3=J$ and $Q_1=Q_2=Q$) version
of (\ref{7dcc}) to make the formula more approachable:
\bea\label{7dccur}
\left( Q^4 + \frac{2N^3}{3}J^3 \right) \left(Q - \frac{N^3}{6} \right)^2
-2 (3 Q^2 + N^3 J) \left(Q^3 - \frac{N^3}{2} J^2 \right) \left(Q - \frac{N^3}{6} \right)
+ \left( Q^3 - \frac{N^3}{2} J^2  \right)^2 &=& 0 ~. \nn\\
\eea

\subsection{The 6d $(2, 0)$ Free Tensor Multiplet}

The charged, rotating AdS$_7$ black holes introduced in the previous subsection
are dual to the 6d $(2, 0)$ theory.
In this subsection we present the free tensor multiplet of the $(2, 0)$ superconformal algebra
needed to construct the single particle partition function.

The 6d $(2, 0)$ superconformal algebra has maximal bosonic subalgebra
$\mathfrak{so}(2,6) \oplus \mathfrak{sp}(4)$, matching the isometry of AdS$_7 \times S^4$.
Local operators in the theory are organized into representations of this subalgebra.
A super-representation of the 6d $(2, 0)$ superconformal algebra is uniquely specified by
the Dynkin labels of its superconformal primary.
For easy comparison with black hole spacetimes, 
we use $SO(6)$ for the Lorentz group and $SO(5)$ for the R-symmetry group, 
instead of $SU(4)$ for Lorentz and $Sp(4)$ for R-symmetry 
used in \cite{Cordova:2016emh}.\footnote{This amounts to the interchanges $j_1 \leftrightarrow j_2$ and $R_1 \leftrightarrow R_2$.}
So we write representations of the bosonic subalgebra as
$$[j_1, j_2, j_3]^{[R_1, R_2]}_E~,$$
where $E$ is the conformal weight, $[j_1, j_2, j_3]$ are the $SO(6)$ Dynkin labels
so that $[1,0,0]$ is the vector ${\bf 6}$,
and $[R_1,R_2]$ are the $SO(5)$ Dynkin labels so that $[1,0]$ is the vector ${\bf 5}$.

The black hole charges used in section \ref{sec:7dBH} refer to the
orthogonal basis that is related to the Dynkin basis as
\bea
&& J_1 = j_1 + \frac{j_2+j_3}{2}~, \qquad
J_2 = \frac{j_2+j_3}{2}~, \qquad
J_3 = \frac{-j_2+j_3}{2}~, \nn\\
&& Q_1 = R_1 + \frac{R_2}{2}~, \qquad ~~~
Q_2 = \frac{R_2}{2}~.
\eea

A (not necessarily the highest) weight $[j_1, j_2, j_3]^{[R_1, R_2]}_E$,
is annihilated by our choice of a supercharge ${\cal Q}$ if it saturates the unitarity bound
\bea\label{7dunitboundmicro}
E &\geq& j_1 + \frac12 j_2 + \frac32 j_3 + 2R_1 + 2R_2 \nn\\
&=& J_1 + J_2 + J_3 + 2Q_1 + 2Q_2~,
\eea
that every weight must satisfy. Such weights correspond to local BPS operators.

\begin{table}\catcode`\-=12
\begin{center}
\begin{tabular}{| c | c | c | c c c c c | c c c c c |}
\hline
& Bosonic Rep. & $E$ & $j_1$ & $j_2$ & $j_3$ & $R_1$ & $R_2$ &
$J_1$ & $J_2$ & $J_3$ & $Q_1$ & $Q_2$ \\
\hline
\hline
\multirow{5}{*}{Free fields} & \multirow{2}{*}{$[0,0,0]^{[1,0]}_2$} & 2 & 0 & 0 & 0 & 1 & 0 & 0 & 0 & 0 & 1 & 0 \\
& & 2 & 0 & 0 & 0 & $-1$ & 2 & 0 & 0 & 0 & 0 & 1 \\
\cline{2-13}
& \multirow{3}{*}{$[0,1,0]^{[0,1]}_\frac52$} & $\frac52$ & 0 & 1 & 0 & 0 & 1 &
$\frac12$ & $\frac12$ & $-\frac12$ & $\frac12$ & $\frac12$ \\
& & $\frac52$ & 1 & $-1$ & 0 & 0 & 1 &
$\frac12$ & $-\frac12$ & $\frac12$ & $\frac12$ & $\frac12$ \\
& & $\frac52$ & $-1$ & 0 & 1 & 0 & 1 &
$-\frac12$ & $\frac12$ & $\frac12$ & $\frac12$ & $\frac12$ \\
\hline
\hline
Eq. of motion & $[0,0,1]^{[0,1]}_\frac72$ & $\frac72$ & 0 & 0 & 1 & 0 & 1 &
$\frac12$ & $\frac12$ & $\frac12$ & $\frac12$ & $\frac12$ \\
\hline
\hline
\multirow{3}{*}{Derivatives} & \multirow{3}{*}{$[1,0,0]^{[0,0]}_1$} & 1 & 1 & 0 & 0 & 0 & 0 & 1 & 0 & 0 & 0 & 0 \\
& & 1 & $-1$ & 1 & 1 & 0 & 0 & 0 & 1 & 0 & 0 & 0 \\
& & 1 & 0 & $-1$ & 1 & 0 & 0 & 0 & 0 & 1 & 0 & 0 \\
\hline
\end{tabular}
\caption{\label{7dfmtable}
Components of the BPS operators in the free tensor multiplet
$D_1 [0,0,0]^{[1,0]}_2$. The first $5$ rows are free fields.
Below the double horizontal lines is the equation of motion,
and the last $3$ rows are derivatives.}
\end{center}
\end{table}

In 6d superconformal theories, a field $[j_1, j_2, j_3]^{[R_1, R_2]}_\Delta$
is free if $j_1 = 0$, at least one of $j_2$ and $j_3$ is zero,
and $E = 2+\frac{j_2+j_3}{2}$.
There is only one multiplet of the 6d $(2, 0)$ superconformal algebra that contains a free field: 
the free tensor multiplet $D_1 [0,0,0]^{[1,0]}_2$ \cite{Cordova:2016emh}.
In Table \ref{7dfmtable} we summarize all weights in the free tensor multiplet that
saturate the unitarity bound (\ref{7dunitboundmicro}).

In Table \ref{7dfmtable}, we have listed $5$ free fields: $2$ scalars and $3$ spinors.
The entry below is an equation of motion that implements a relation between two spinors, so it can be counted as a negative field.
The three last entries are derivatives that may act on any of the fields
and on the equation of motion, to produce their BPS descendants.
The gradient in $6$ dimensions has $6$ components but only $3$ preserve the BPS-ness of the field. 
The $5$ free fields, modulo the equation of motion, and with possible derivatives taken into account,
are the exhaustive list of supersymmetric operators in the free tensor multiplet.

\subsection{The Supersymmetric Ensemble}

We now compute the single particle BPS partition function as a trace over the free tensor multiplet states given in Table \ref{7dfmtable}, with fugacities $(p,q,r,x,y)$ conjugate to each charge:
\bea
Z_{\rm sp} &\equiv& {\rm Tr}_{\rm BPS} \left[ p^{2J_1} q^{2J_2} r^{2J_3} x^{2Q_1} y^{2Q_2} \right]\cr
%
%
\label{7dsppf}
&=& \frac{x^2+y^2+xypqr \left( \frac{1}{p^2} + \frac{1}{q^2} + \frac{1}{r^2} - 1\right)
}{(1-p^2)(1-q^2)(1-r^2)}~.
\eea
The $-1$ inside the parenthesis in the numerator is due to the equation of motion. The 
geometric series in $p^2$, $q^2$, and $r^2$ are from the derivatives. 
(\ref{7dsppf}) is the single particle partition function.

The grand canonical partition function over the full Hilbert space
is given by the ordinary exponential of the single particle partition function:
$Z \equiv \exp[Z_{\rm sp}] $. We use it to compute the macroscopic charges as statistical averages:
\bea\label{7dchgpotrel}
Q_1 &=&  \frac{2x^2+xypqr \left( \frac{1}{p^2} + \frac{1}{q^2} + \frac{1}{r^2} - 1\right)
}{2(1-p^2)(1-q^2)(1-r^2)}~, \nn\\
J_1 &=&  \frac{2p^2(x^2+y^2)+xypqr (1+p^2)\left( -\frac{1}{p^2} + \frac{1}{q^2} + \frac{1}{r^2} - 1\right)
+ 4 xypqr}{2(1-p^2)^2(1-q^2)(1-r^2)}~. 
\eea
Analogous expressions for $Q_2$, $J_2$ and $J_3$ follow by permutations of indices. 

Finally, we define the supersymmetric ensemble as a grand canonical ensemble
where the operators that are separated in the charge space along the direction
of the preserved supercharge are weighed equally.
The preserved supercharge ${\cal Q}$ carries quantum numbers
$(E, J_1, J_2, J_3, Q_1, Q_2) =
\left( \frac12, -\frac12, -\frac12, -\frac12, \frac12, \frac12 \right)$,
so the supersymmetric ensemble corresponds to imposing the relation
\bea\label{7dsusyens}
\frac{xy}{pqr} &=& 1~,
\eea
between the fugacities.

In the supersymmetric ensemble defined by (\ref{7dsusyens}), there is one relation between the $5$ charges (\ref{7dchgpotrel}):
\bea\label{7dccmicro}
&& \frac12 (J_1+J_2+J_3) + \frac12 (Q_1^2 + Q_2^2 + 4Q_1Q_2)
- \frac{Q_1Q_2(Q_1+Q_2) - \frac12 (J_1J_2 + J_2J_3 + J_3J_1)}
{Q_1+Q_2- \frac12} \nn\\
&=& \sqrt{
\left( \frac12 (J_1+J_2+J_3) + \frac12 (Q_1^2 + Q_2^2 + 4Q_1Q_2) \right)^2
- \left( J_1J_2J_3 + Q_1^2Q_2^2 \right)}~.
\eea
It is precisely the supersymmetric AdS$_7$ black hole charge constraint (\ref{7dcc}) with the numerical values
\bea\label{7dNvalue}
\frac{2}{3}N^3 = 1 \quad \leftrightarrow \quad   \frac{\pi^2}{8G_7} = 1~.
\eea
%
%
%
We interpret this relative scale of all charges as the effective number of free multiplets needed to account for the constraint.
It is satisfying that the numerical factor $\frac{2}{3}<1$ since the interpolation from weak to strong coupling is expected to decrease the effective number of degrees of freedom.

The formulae simplify significantly when we do not distinguish between the $3$ angular momenta and between the $2$ electric charges. 
Let $x=y$ and $p=q=r$ in (\ref{7dchgpotrel})):
\bea\label{7dchgpotrelur}
Q &=&  \frac{2x^2+ 3x^2p - x^2p^3}{2(1-p^2)^3}~, \nn\\
J &=&  \frac{x^2p + 4x^2p^2  + 4 x^2p^3 -x^2p^5}{2(1-p^2)^4}~.
\eea
The definition of the supersymmetric ensemble \eqref{7dsusyens} simplifies to $x^2=p^3$, 
and then the charges \eqref{7dchgpotrelur} satisfy
the unrefined charge constraint (\ref{7dccur}) with $\frac23 N^3 = 1$.

\section{Discussion}
\label{sec:discussion}

In the body of this paper, we derived the supersymmetric charge constraint for the AdS$_{4,5,7}$ black holes using the simple prescription given in section \ref{sec:gendim}.
We think the computations are illuminating, especially because they are so simple.
However, we acknowledge that, in its current form, the argument is heuristic and subject to significant concerns. 
These challenges are the subject of this final section. We divide them into four issues, even though their possible resolutions are interrelated:
\begin{enumerate}
\item[1)] {\it Coupling dependence.} Unlike the index, the partition function depends on the coupling $g_\mathrm{YM}$. We study an ensemble of states generated by free fields and, even so, we compare the result to black holes that correspond to strong coupling. 

\item[2)] {\it Gauge dynamics.} In each case, we consider a single free field, rather than the dynamics due to gauge degrees of freedom. 

The dependence on Newton's constant is determined by dimensional analysis in gravity,
while the dependence on the rank in the dual CFT is reproduced by assuming that it is in its deconfined phase.
However, a numerical constant of ${\cal O}(1)$ is put in by hand. 

\item[3)] {\it Classical statistics.} We consider a classical gas of BPS particles. Technically, we take the multiparticle partition function to be the simple exponential of the single particle partition function, rather than the plethystic exponential. We did not justify why this approximation is sufficient. 

\item[4)] {\it The supersymmetric ensemble}: is defined so states that differ by the charges of the preserved supercharge ${\cal Q}$ are given equal weight. This is motivated by the real part of the supersymmetry constraint on complex fugacities, which are well established in supersymmetric black hole spacetimes \cite{Hosseini:2017mds,Cabo-Bizet:2018ehj,Choi:2018hmj,Larsen:2019oll,Larsen:2021wnu}. We did not provide a self-contained justification of this ensemble in the CFT, except for the CFT$_2$ argument presented in section \ref{sec:3d}. 
\end{enumerate}

The numerical factor mentioned in 2) presents a concrete goal that involves several of these issues.
If all fields were genuinely free, the number of independent multiplets would be 
$N^2$ in AdS$_5$/CFT$_4$, from the dimension of the $SU(N)$ gauge group of ${\cal N}=4$ SYM,
and similarly in other dimensions.
This type of a na\"{\i}ve count of multiplets would not even take the projection onto gauge singlets
for the physical Hilbert space into account.
This can in principle be addressed by upgrading to a matrix model and, in particular,
confronting 3) \cite{Aharony:2003sx}.
However, this still leaves 1), the dependence on the coupling constant:
some of the BPS states in the free theory may gain anomalous dimensions and be lifted from being BPS.
The projection onto singlets and the dependence on the coupling both suggest that
the na\"{\i}ve scaling in $N$ overcounts the microscopic states rather than undercounts.
It is therefore encouraging that all the needed ${\cal O}(1)$ adjustments are smaller than $1$.
Table \ref{summary} records the rescaling factors $\frac{\sqrt{2}}{3} N^{3/2}$, $\frac12 N^2$
and $\frac23 N^3$ for the AdS$_4$, AdS$_5$, and AdS$_7$ charge constraints, respectively.
The situation is reminiscent of the famous $3/4$-renormalization of high temperature
D3-brane entropy as the coupling is taken from weak to strong \cite{Gubser:1996de}.

Our discussion of supersymmetric black holes in AdS$_3$
is on more solid footing than in the higher dimensions. 
That is because the superalgebra is much stronger,
it gives a complete basis of characters for both short and long supermultiplets of the $\mathcal{N} = 4$ super-Virasoro algebra, and so no free field assumption is needed.
In this context the supersymmetric ensemble is justified by a symmetry,
and the constraint we find agrees precisely with the black hole side,
with no numerical factor put in by hand.
These results offer a template for higher dimensions that we have pursued,
especially when addressing 4), but it is possible that other lessons remain hidden in plain sight. 

Our approach is fundamentally limited by us studying the partition function, rather than the supersymmetric index. Therefore, our computation is unavoidably subject to dependence on the coupling constant that is beyond our control. On the other hand, although the index is an invaluable tool for circumventing the coupling dependence, it has its own structural limitations. Because it is insensitive to many quantum states, it can at best provide a lower bound on the black hole entropy, and so any agreement is only genuinely successful if it is understood why cancellations are subleading. The limitations of the index are especially pertinent in our context, the constraint on charges that is satisfied by all supersymmetric black holes in AdS spacetimes. That is because the index is independent of the relevant physical variable, to the best of our understanding. 

For the future, the vision ultimately is that all the various contributions to the partition function, in gravity and in CFT, whether boundary conditions correspond to an index or not, can be disentangled.
Significant strides have been taken towards this goal in the most favorable circumstances,
such as asymptotically flat spacetimes with at least $\frac{1}{8}$ of the supersymmetries \cite{Sen:2009vz,Sen:2009gy,Sen:2014aja,Iliesiu:2022kny,Iliesiu:2022onk}.
For asymptotically AdS spacetimes with maximal supersymmetry, the setting we have studied,
the current research frontier is at a lower level of understanding,
but recent years have witnessed much progress, using a variety of techniques \cite{Iliesiu:2020qvm,Heydeman:2020hhw,Boruch:2022tno}.
The work presented in this article, including the challenges discussed in this section,
is a contribution to these developments. 

\section*{Acknowledgements}

We would like to thank Sunjin Choi for helpful discussions.
FL thanks Stanford ITP for hospitality while this article was completed.
SL is supported by the Rackham Predoctoral Fellowship.
This work was supported in part by DoE grant DE-SC0007859.

\bibliography{AdSccdraftv4}{}

\providecommand{\href}[2]{#2}\begingroup\raggedright\begin{thebibliography}{10}

\bibitem{Cvetic:2005zi}
M.~Cvetic, G.W.~Gibbons, H.~Lu and C.N.~Pope, \emph{{Rotating black holes in
  gauged supergravities: Thermodynamics, supersymmetric limits, topological
  solitons and time machines}},
  \href{https://arxiv.org/abs/hep-th/0504080}{{\ttfamily hep-th/0504080}}.

\bibitem{Cassani:2019mms}
D.~Cassani and L.~Papini, \emph{{The BPS limit of rotating AdS black hole
  thermodynamics}}, \href{https://doi.org/10.1007/JHEP09(2019)079}{\emph{JHEP}
  {\bfseries 09} (2019) 079}
  [\href{https://arxiv.org/abs/1906.10148}{{\ttfamily 1906.10148}}].

\bibitem{Banados:1992wn}
M.~Banados, C.~Teitelboim and J.~Zanelli, \emph{{The Black hole in
  three-dimensional space-time}},
  \href{https://doi.org/10.1103/PhysRevLett.69.1849}{\emph{Phys. Rev. Lett.}
  {\bfseries 69} (1992) 1849}
  [\href{https://arxiv.org/abs/hep-th/9204099}{{\ttfamily hep-th/9204099}}].

\bibitem{Banados:1992gq}
M.~Banados, M.~Henneaux, C.~Teitelboim and J.~Zanelli, \emph{{Geometry of the
  (2+1) black hole}},
  \href{https://doi.org/10.1103/PhysRevD.48.1506}{\emph{Phys. Rev. D}
  {\bfseries 48} (1993) 1506}
  [\href{https://arxiv.org/abs/gr-qc/9302012}{{\ttfamily gr-qc/9302012}}].

\bibitem{Chong:2004na}
Z.W.~Chong, M.~Cvetic, H.~Lu and C.N.~Pope, \emph{{Charged rotating black holes
  in four-dimensional gauged and ungauged supergravities}},
  \href{https://doi.org/10.1016/j.nuclphysb.2005.03.034}{\emph{Nucl. Phys. B}
  {\bfseries 717} (2005) 246}
  [\href{https://arxiv.org/abs/hep-th/0411045}{{\ttfamily hep-th/0411045}}].

\bibitem{Hristov:2019mqp}
K.~Hristov, S.~Katmadas and C.~Toldo, \emph{{Matter-coupled supersymmetric
  Kerr-Newman-AdS$_4$ black holes}},
  \href{https://doi.org/10.1103/PhysRevD.100.066016}{\emph{Phys. Rev. D}
  {\bfseries 100} (2019) 066016}
  [\href{https://arxiv.org/abs/1907.05192}{{\ttfamily 1907.05192}}].

\bibitem{Gutowski:2004ez}
J.B.~Gutowski and H.S.~Reall, \emph{{Supersymmetric AdS(5) black holes}},
  \href{https://doi.org/10.1088/1126-6708/2004/02/006}{\emph{JHEP} {\bfseries
  02} (2004) 006} [\href{https://arxiv.org/abs/hep-th/0401042}{{\ttfamily
  hep-th/0401042}}].

\bibitem{Gutowski:2004yv}
J.B.~Gutowski and H.S.~Reall, \emph{{General supersymmetric AdS(5) black
  holes}}, \href{https://doi.org/10.1088/1126-6708/2004/04/048}{\emph{JHEP}
  {\bfseries 04} (2004) 048}
  [\href{https://arxiv.org/abs/hep-th/0401129}{{\ttfamily hep-th/0401129}}].

\bibitem{Chong:2005da}
Z.~Chong, M.~Cvetic, H.~Lu and C.~Pope, \emph{{Five-dimensional gauged
  supergravity black holes with independent rotation parameters}},
  \href{https://doi.org/10.1103/PhysRevD.72.041901}{\emph{Phys. Rev. D}
  {\bfseries 72} (2005) 041901}
  [\href{https://arxiv.org/abs/hep-th/0505112}{{\ttfamily hep-th/0505112}}].

\bibitem{Chong:2005hr}
Z.-W.~Chong, M.~Cvetic, H.~Lu and C.~Pope, \emph{{General non-extremal rotating
  black holes in minimal five-dimensional gauged supergravity}},
  \href{https://doi.org/10.1103/PhysRevLett.95.161301}{\emph{Phys. Rev. Lett.}
  {\bfseries 95} (2005) 161301}
  [\href{https://arxiv.org/abs/hep-th/0506029}{{\ttfamily hep-th/0506029}}].

\bibitem{Kunduri:2006ek}
H.K.~Kunduri, J.~Lucietti and H.S.~Reall, \emph{{Supersymmetric multi-charge
  AdS(5) black holes}},
  \href{https://doi.org/10.1088/1126-6708/2006/04/036}{\emph{JHEP} {\bfseries
  04} (2006) 036} [\href{https://arxiv.org/abs/hep-th/0601156}{{\ttfamily
  hep-th/0601156}}].

\bibitem{Wu:2011gq}
S.-Q.~Wu, \emph{{General Nonextremal Rotating Charged AdS Black Holes in
  Five-dimensional $U(1)^3$ Gauged Supergravity: A Simple Construction
  Method}}, \href{https://doi.org/10.1016/j.physletb.2011.12.031}{\emph{Phys.
  Lett. B} {\bfseries 707} (2012) 286}
  [\href{https://arxiv.org/abs/1108.4159}{{\ttfamily 1108.4159}}].

\bibitem{Chong:2004dy}
Z.W.~Chong, M.~Cvetic, H.~Lu and C.N.~Pope, \emph{{Non-extremal charged
  rotating black holes in seven-dimensional gauged supergravity}},
  \href{https://doi.org/10.1016/j.physletb.2005.07.054}{\emph{Phys. Lett. B}
  {\bfseries 626} (2005) 215}
  [\href{https://arxiv.org/abs/hep-th/0412094}{{\ttfamily hep-th/0412094}}].

\bibitem{Chow:2007ts}
D.D.K.~Chow, \emph{{Equal charge black holes and seven dimensional gauged
  supergravity}},
  \href{https://doi.org/10.1088/0264-9381/25/17/175010}{\emph{Class. Quant.
  Grav.} {\bfseries 25} (2008) 175010}
  [\href{https://arxiv.org/abs/0711.1975}{{\ttfamily 0711.1975}}].

\bibitem{Wu:2011gp}
S.-Q.~Wu, \emph{{Two-charged non-extremal rotating black holes in
  seven-dimensional gauged supergravity: The Single-rotation case}},
  \href{https://doi.org/10.1016/j.physletb.2011.10.026}{\emph{Phys. Lett. B}
  {\bfseries 705} (2011) 383}
  [\href{https://arxiv.org/abs/1108.4158}{{\ttfamily 1108.4158}}].

\bibitem{Chow:2011fh}
D.D.K.~Chow, \emph{{Single-rotation two-charge black holes in gauged
  supergravity}},  \href{https://arxiv.org/abs/1108.5139}{{\ttfamily
  1108.5139}}.

\bibitem{Bobev:2023bxl}
N.~Bobev, M.~David, J.~Hong and R.~Mouland, \emph{{AdS$_{7}$ black holes from
  rotating M5-branes}},
  \href{https://doi.org/10.1007/JHEP09(2023)143}{\emph{JHEP} {\bfseries 09}
  (2023) 143} [\href{https://arxiv.org/abs/2307.06364}{{\ttfamily
  2307.06364}}].

\bibitem{Maldacena:1997re}
J.M.~Maldacena, \emph{{The Large N limit of superconformal field theories and
  supergravity}}, \href{https://doi.org/10.4310/ATMP.1998.v2.n2.a1}{\emph{Adv.
  Theor. Math. Phys.} {\bfseries 2} (1998) 231}
  [\href{https://arxiv.org/abs/hep-th/9711200}{{\ttfamily hep-th/9711200}}].

\bibitem{Gubser:1998bc}
S.S.~Gubser, I.R.~Klebanov and A.M.~Polyakov, \emph{{Gauge theory correlators
  from noncritical string theory}},
  \href{https://doi.org/10.1016/S0370-2693(98)00377-3}{\emph{Phys. Lett. B}
  {\bfseries 428} (1998) 105}
  [\href{https://arxiv.org/abs/hep-th/9802109}{{\ttfamily hep-th/9802109}}].

\bibitem{Witten:1998qj}
E.~Witten, \emph{{Anti-de Sitter space and holography}},
  \href{https://doi.org/10.4310/ATMP.1998.v2.n2.a2}{\emph{Adv. Theor. Math.
  Phys.} {\bfseries 2} (1998) 253}
  [\href{https://arxiv.org/abs/hep-th/9802150}{{\ttfamily hep-th/9802150}}].

\bibitem{Benini:2015eyy}
F.~Benini, K.~Hristov and A.~Zaffaroni, \emph{{Black hole microstates in
  AdS$_{4}$ from supersymmetric localization}},
  \href{https://doi.org/10.1007/JHEP05(2016)054}{\emph{JHEP} {\bfseries 05}
  (2016) 054} [\href{https://arxiv.org/abs/1511.04085}{{\ttfamily
  1511.04085}}].

\bibitem{Benini:2016rke}
F.~Benini, K.~Hristov and A.~Zaffaroni, \emph{{Exact microstate counting for
  dyonic black holes in AdS4}},
  \href{https://doi.org/10.1016/j.physletb.2017.05.076}{\emph{Phys. Lett. B}
  {\bfseries 771} (2017) 462}
  [\href{https://arxiv.org/abs/1608.07294}{{\ttfamily 1608.07294}}].

\bibitem{Hosseini:2017mds}
S.M.~Hosseini, K.~Hristov and A.~Zaffaroni, \emph{{An extremization principle
  for the entropy of rotating BPS black holes in AdS$_{5}$}},
  \href{https://doi.org/10.1007/JHEP07(2017)106}{\emph{JHEP} {\bfseries 07}
  (2017) 106} [\href{https://arxiv.org/abs/1705.05383}{{\ttfamily
  1705.05383}}].

\bibitem{Azzurli:2017kxo}
F.~Azzurli, N.~Bobev, P.M.~Crichigno, V.S.~Min and A.~Zaffaroni, \emph{{A
  universal counting of black hole microstates in AdS$_{4}$}},
  \href{https://doi.org/10.1007/JHEP02(2018)054}{\emph{JHEP} {\bfseries 02}
  (2018) 054} [\href{https://arxiv.org/abs/1707.04257}{{\ttfamily
  1707.04257}}].

\bibitem{Cabo-Bizet:2018ehj}
A.~Cabo-Bizet, D.~Cassani, D.~Martelli and S.~Murthy, \emph{{Microscopic origin
  of the Bekenstein-Hawking entropy of supersymmetric AdS$_{5}$ black holes}},
  \href{https://doi.org/10.1007/JHEP10(2019)062}{\emph{JHEP} {\bfseries 10}
  (2019) 062} [\href{https://arxiv.org/abs/1810.11442}{{\ttfamily
  1810.11442}}].

\bibitem{Choi:2018hmj}
S.~Choi, J.~Kim, S.~Kim and J.~Nahmgoong, \emph{{Large AdS black holes from
  QFT}},  \href{https://arxiv.org/abs/1810.12067}{{\ttfamily 1810.12067}}.

\bibitem{Benini:2018ywd}
F.~Benini and P.~Milan, \emph{{Black Holes in 4D $\mathcal{N}$=4
  Super-Yang-Mills Field Theory}},
  \href{https://doi.org/10.1103/PhysRevX.10.021037}{\emph{Phys. Rev. X}
  {\bfseries 10} (2020) 021037}
  [\href{https://arxiv.org/abs/1812.09613}{{\ttfamily 1812.09613}}].

\bibitem{Choi:2018fdc}
S.~Choi, C.~Hwang, S.~Kim and J.~Nahmgoong, \emph{{Entropy Functions of BPS
  Black Holes in AdS$_{4}$ and AdS$_{6}$}},
  \href{https://doi.org/10.3938/jkps.76.101}{\emph{J. Korean Phys. Soc.}
  {\bfseries 76} (2020) 101}
  [\href{https://arxiv.org/abs/1811.02158}{{\ttfamily 1811.02158}}].

\bibitem{Choi:2019miv}
S.~Choi and S.~Kim, \emph{{Large AdS$_6$ black holes from CFT$_5$}},
  \href{https://arxiv.org/abs/1904.01164}{{\ttfamily 1904.01164}}.

\bibitem{Choi:2019zpz}
S.~Choi, C.~Hwang and S.~Kim, \emph{{Quantum vortices, M2-branes and black
  holes}},  \href{https://arxiv.org/abs/1908.02470}{{\ttfamily 1908.02470}}.

\bibitem{Nian:2019pxj}
J.~Nian and L.A.~Pando~Zayas, \emph{{Microscopic entropy of rotating
  electrically charged AdS$_{4}$ black holes from field theory localization}},
  \href{https://doi.org/10.1007/JHEP03(2020)081}{\emph{JHEP} {\bfseries 03}
  (2020) 081} [\href{https://arxiv.org/abs/1909.07943}{{\ttfamily
  1909.07943}}].

\bibitem{Hosseini:2019iad}
S.M.~Hosseini, K.~Hristov and A.~Zaffaroni, \emph{{Gluing gravitational blocks
  for AdS black holes}},
  \href{https://doi.org/10.1007/JHEP12(2019)168}{\emph{JHEP} {\bfseries 12}
  (2019) 168} [\href{https://arxiv.org/abs/1909.10550}{{\ttfamily
  1909.10550}}].

\bibitem{Benini:2019dyp}
F.~Benini, D.~Gang and L.A.~Pando~Zayas, \emph{{Rotating Black Hole Entropy
  from M5 Branes}}, \href{https://doi.org/10.1007/JHEP03(2020)057}{\emph{JHEP}
  {\bfseries 03} (2020) 057}
  [\href{https://arxiv.org/abs/1909.11612}{{\ttfamily 1909.11612}}].

\bibitem{Choi:2019dfu}
S.~Choi and C.~Hwang, \emph{{Universal 3d Cardy Block and Black Hole Entropy}},
  \href{https://doi.org/10.1007/JHEP03(2020)068}{\emph{JHEP} {\bfseries 03}
  (2020) 068} [\href{https://arxiv.org/abs/1911.01448}{{\ttfamily
  1911.01448}}].

\bibitem{Cabo-Bizet:2020nkr}
A.~Cabo-Bizet, D.~Cassani, D.~Martelli and S.~Murthy, \emph{{The large-$N$
  limit of the 4d $ \mathcal{N} $ = 1 superconformal index}},
  \href{https://doi.org/10.1007/JHEP11(2020)150}{\emph{JHEP} {\bfseries 11}
  (2020) 150} [\href{https://arxiv.org/abs/2005.10654}{{\ttfamily
  2005.10654}}].

\bibitem{Larsen:2021wnu}
F.~Larsen and S.~Lee, \emph{{Microscopic entropy of AdS$_{3}$ black holes
  revisited}}, \href{https://doi.org/10.1007/JHEP07(2021)038}{\emph{JHEP}
  {\bfseries 07} (2021) 038}
  [\href{https://arxiv.org/abs/2101.08497}{{\ttfamily 2101.08497}}].

\bibitem{Choi:2021rxi}
S.~Choi, S.~Jeong, S.~Kim and E.~Lee, \emph{{Exact QFT duals of AdS black
  holes}}, \href{https://doi.org/10.1007/JHEP09(2023)138}{\emph{JHEP}
  {\bfseries 09} (2023) 138}
  [\href{https://arxiv.org/abs/2111.10720}{{\ttfamily 2111.10720}}].

\bibitem{Aharony:2024ntg}
O.~Aharony, O.~Mamroud, S.~Nowik and M.~Weissman, \emph{{The Bethe Ansatz for
  the superconformal index with unequal angular momenta}},
  \href{https://arxiv.org/abs/2402.03977}{{\ttfamily 2402.03977}}.

\bibitem{Kinney:2005ej}
J.~Kinney, J.M.~Maldacena, S.~Minwalla and S.~Raju, \emph{{An Index for 4
  dimensional super conformal theories}},
  \href{https://doi.org/10.1007/s00220-007-0258-7}{\emph{Commun. Math. Phys.}
  {\bfseries 275} (2007) 209}
  [\href{https://arxiv.org/abs/hep-th/0510251}{{\ttfamily hep-th/0510251}}].

\bibitem{Romelsberger:2005eg}
C.~Romelsberger, \emph{{Counting chiral primaries in N = 1, d=4 superconformal
  field theories}},
  \href{https://doi.org/10.1016/j.nuclphysb.2006.03.037}{\emph{Nucl. Phys. B}
  {\bfseries 747} (2006) 329}
  [\href{https://arxiv.org/abs/hep-th/0510060}{{\ttfamily hep-th/0510060}}].

\bibitem{Bhattacharya:2008zy}
J.~Bhattacharya, S.~Bhattacharyya, S.~Minwalla and S.~Raju, \emph{{Indices for
  Superconformal Field Theories in 3,5 and 6 Dimensions}},
  \href{https://doi.org/10.1088/1126-6708/2008/02/064}{\emph{JHEP} {\bfseries
  02} (2008) 064} [\href{https://arxiv.org/abs/0801.1435}{{\ttfamily
  0801.1435}}].

\bibitem{Chang:2022mjp}
C.-M.~Chang and Y.-H.~Lin, \emph{{Words to describe a black hole}},
  \href{https://doi.org/10.1007/JHEP02(2023)109}{\emph{JHEP} {\bfseries 02}
  (2023) 109} [\href{https://arxiv.org/abs/2209.06728}{{\ttfamily
  2209.06728}}].

\bibitem{Choi:2022caq}
S.~Choi, S.~Kim, E.~Lee and J.~Park, \emph{{The shape of non-graviton operators
  for $SU(2)$}},  \href{https://arxiv.org/abs/2209.12696}{{\ttfamily
  2209.12696}}.

\bibitem{Choi:2023znd}
S.~Choi, S.~Kim, E.~Lee, S.~Lee and J.~Park, \emph{{Towards quantum black hole
  microstates}}, \href{https://doi.org/10.1007/JHEP11(2023)175}{\emph{JHEP}
  {\bfseries 11} (2023) 175}
  [\href{https://arxiv.org/abs/2304.10155}{{\ttfamily 2304.10155}}].

\bibitem{Budzik:2023vtr}
K.~Budzik, H.~Murali and P.~Vieira, \emph{{Following Black Hole States}},
  \href{https://arxiv.org/abs/2306.04693}{{\ttfamily 2306.04693}}.

\bibitem{Choi:2023vdm}
J.~Choi, S.~Choi, S.~Kim, J.~Lee and S.~Lee, \emph{{Finite $N$ black hole
  cohomologies}},  \href{https://arxiv.org/abs/2312.16443}{{\ttfamily
  2312.16443}}.

\bibitem{Cordova:2016emh}
C.~Cordova, T.T.~Dumitrescu and K.~Intriligator, \emph{{Multiplets of
  Superconformal Symmetry in Diverse Dimensions}},
  \href{https://doi.org/10.1007/JHEP03(2019)163}{\emph{JHEP} {\bfseries 03}
  (2019) 163} [\href{https://arxiv.org/abs/1612.00809}{{\ttfamily
  1612.00809}}].

\bibitem{Brown:1986nw}
J.~Brown and M.~Henneaux, \emph{{Central Charges in the Canonical Realization
  of Asymptotic Symmetries: An Example from Three-Dimensional Gravity}},
  \href{https://doi.org/10.1007/BF01211590}{\emph{Commun. Math. Phys.}
  {\bfseries 104} (1986) 207}.

\bibitem{Eguchi:1987sm}
T.~Eguchi and A.~Taormina, \emph{{Unitary Representations of $N=4$
  Superconformal Algebra}},
  \href{https://doi.org/10.1016/0370-2693(87)91679-0}{\emph{Phys. Lett. B}
  {\bfseries 196} (1987) 75}.

\bibitem{Eguchi:1987wf}
T.~Eguchi and A.~Taormina, \emph{{Character Formulas for the $N=4$
  Superconformal Algebra}},
  \href{https://doi.org/10.1016/0370-2693(88)90778-2}{\emph{Phys. Lett. B}
  {\bfseries 200} (1988) 315}.

\bibitem{Sevrin:1988ew}
A.~Sevrin, W.~Troost and A.~Van~Proeyen, \emph{{Superconformal Algebras in
  Two-Dimensions with N=4}},
  \href{https://doi.org/10.1016/0370-2693(88)90645-4}{\emph{Phys. Lett.}
  {\bfseries B208} (1988) 447}.

\bibitem{Eguchi:1988af}
T.~Eguchi and A.~Taormina, \emph{{On the Unitary Representations of $N=2$ and
  $N=4$ Superconformal Algebras}},
  \href{https://doi.org/10.1016/0370-2693(88)90360-7}{\emph{Phys. Lett. B}
  {\bfseries 210} (1988) 125}.

\bibitem{Kim:2006he}
S.~Kim and K.-M.~Lee, \emph{{1/16-BPS Black Holes and Giant Gravitons in the
  AdS(5) X S**5 Space}},
  \href{https://doi.org/10.1088/1126-6708/2006/12/077}{\emph{JHEP} {\bfseries
  12} (2006) 077} [\href{https://arxiv.org/abs/hep-th/0607085}{{\ttfamily
  hep-th/0607085}}].

\bibitem{Larsen:2019oll}
F.~Larsen, J.~Nian and Y.~Zeng, \emph{{AdS$_{5}$ black hole entropy near the
  BPS limit}}, \href{https://doi.org/10.1007/JHEP06(2020)001}{\emph{JHEP}
  {\bfseries 06} (2020) 001}
  [\href{https://arxiv.org/abs/1907.02505}{{\ttfamily 1907.02505}}].

\bibitem{Cassani:2021fyv}
D.~Cassani and Z.~Komargodski, \emph{{EFT and the SUSY Index on the 2nd
  Sheet}}, \href{https://doi.org/10.21468/SciPostPhys.11.1.004}{\emph{SciPost
  Phys.} {\bfseries 11} (2021) 004}
  [\href{https://arxiv.org/abs/2104.01464}{{\ttfamily 2104.01464}}].

\bibitem{Larsen:2020lhg}
F.~Larsen and S.~Paranjape, \emph{{Thermodynamics of near BPS black holes in
  AdS$_{4}$ and AdS$_{7}$}},
  \href{https://doi.org/10.1007/JHEP10(2021)198}{\emph{JHEP} {\bfseries 10}
  (2021) 198} [\href{https://arxiv.org/abs/2010.04359}{{\ttfamily
  2010.04359}}].

\bibitem{Aharony:2003sx}
O.~Aharony, J.~Marsano, S.~Minwalla, K.~Papadodimas and M.~Van~Raamsdonk,
  \emph{{The Hagedorn - deconfinement phase transition in weakly coupled large
  N gauge theories}},
  \href{https://doi.org/10.4310/ATMP.2004.v8.n4.a1}{\emph{Adv. Theor. Math.
  Phys.} {\bfseries 8} (2004) 603}
  [\href{https://arxiv.org/abs/hep-th/0310285}{{\ttfamily hep-th/0310285}}].

\bibitem{Gubser:1996de}
S.S.~Gubser, I.R.~Klebanov and A.W.~Peet, \emph{{Entropy and temperature of
  black 3-branes}}, \href{https://doi.org/10.1103/PhysRevD.54.3915}{\emph{Phys.
  Rev. D} {\bfseries 54} (1996) 3915}
  [\href{https://arxiv.org/abs/hep-th/9602135}{{\ttfamily hep-th/9602135}}].

\bibitem{Sen:2009vz}
A.~Sen, \emph{{Arithmetic of Quantum Entropy Function}},
  \href{https://doi.org/10.1088/1126-6708/2009/08/068}{\emph{JHEP} {\bfseries
  08} (2009) 068} [\href{https://arxiv.org/abs/0903.1477}{{\ttfamily
  0903.1477}}].

\bibitem{Sen:2009gy}
A.~Sen, \emph{{Arithmetic of N=8 Black Holes}},
  \href{https://doi.org/10.1007/JHEP02(2010)090}{\emph{JHEP} {\bfseries 02}
  (2010) 090} [\href{https://arxiv.org/abs/0908.0039}{{\ttfamily 0908.0039}}].

\bibitem{Sen:2014aja}
A.~Sen, \emph{{Microscopic and Macroscopic Entropy of Extremal Black Holes in
  String Theory}}, \href{https://doi.org/10.1007/s10714-014-1711-5}{\emph{Gen.
  Rel. Grav.} {\bfseries 46} (2014) 1711}
  [\href{https://arxiv.org/abs/1402.0109}{{\ttfamily 1402.0109}}].

\bibitem{Iliesiu:2022kny}
L.V.~Iliesiu, S.~Murthy and G.J.~Turiaci, \emph{{Black hole microstate counting
  from the gravitational path integral}},
  \href{https://arxiv.org/abs/2209.13602}{{\ttfamily 2209.13602}}.

\bibitem{Iliesiu:2022onk}
L.V.~Iliesiu, S.~Murthy and G.J.~Turiaci, \emph{{Revisiting the Logarithmic
  Corrections to the Black Hole Entropy}},
  \href{https://arxiv.org/abs/2209.13608}{{\ttfamily 2209.13608}}.

\bibitem{Iliesiu:2020qvm}
L.V.~Iliesiu and G.J.~Turiaci, \emph{{The statistical mechanics of
  near-extremal black holes}},
  \href{https://doi.org/10.1007/JHEP05(2021)145}{\emph{JHEP} {\bfseries 05}
  (2021) 145} [\href{https://arxiv.org/abs/2003.02860}{{\ttfamily
  2003.02860}}].

\bibitem{Heydeman:2020hhw}
M.~Heydeman, L.V.~Iliesiu, G.J.~Turiaci and W.~Zhao, \emph{{The statistical
  mechanics of near-BPS black holes}},
  \href{https://doi.org/10.1088/1751-8121/ac3be9}{\emph{J. Phys. A} {\bfseries
  55} (2022) 014004} [\href{https://arxiv.org/abs/2011.01953}{{\ttfamily
  2011.01953}}].

\bibitem{Boruch:2022tno}
J.~Boruch, M.T.~Heydeman, L.V.~Iliesiu and G.J.~Turiaci, \emph{{BPS and
  near-BPS black holes in $AdS_5$ and their spectrum in $\mathcal{N}=4$ SYM}},
  \href{https://arxiv.org/abs/2203.01331}{{\ttfamily 2203.01331}}.

\end{thebibliography}\endgroup
\bibliographystyle{JHEP}

\end{document}